\documentclass[aps,prd,groupedaddress,amsmath,amssymb,amsfonts,nofootinbib,preprint,superscriptaddress]{revtex4-1}
\usepackage{amssymb,amsfonts}
\usepackage{mathrsfs}

\usepackage{tikz}
\usetikzlibrary{patterns}
\usetikzlibrary{decorations.pathreplacing}
\usetikzlibrary{calc}

\usepackage{amsthm}

\usepackage[colorlinks, allcolors=blue!70!black, linktocpage]{hyperref}
\usepackage{accents}
\usepackage{graphicx}
\usepackage[utf8x]{inputenc}
\usepackage{float}
\usepackage{amsmath}
\DeclareMathAlphabet{\mathpzc}{OT1}{pzc}{m}{it}

\usepackage{titlesec}
\usepackage{cases}
\usepackage{mathtools}
\usepackage{color}
\usepackage{cleveref}
\usepackage{braket,mathtools,float,amsfonts}
\usepackage{comment}
\usepackage{extarrows}

\newcommand{\be}{\begin{equation}}
\newcommand{\ee}{\end{equation}}
\newcommand{\bea}{\begin{eqnarray}}
\newcommand{\eea}{\end{eqnarray}}

\definecolor{Gray}{gray}{0.95}
\definecolor{LightCyan}{rgb}{0.88,1,1}

\def\Tr{{\text{Tr}}}

\newcommand{\defn}{\mathrel{\mathop:}=} 

\crefname{lem}{lemma}{lemmas}
\crefname{thm}{theorem}{theorems}
\crefname{cor}{corollary}{corollaries}
\crefname{rem}{remark}{remarks}
\crefname{prop}{proposition}{propositions}

\newcommand{\darkred}{red!85!black}
\newcommand{\darkblue}{cyan!85!black}
\newcommand{\lightblue}{cyan!10!white}
\newcommand{\rt}{-sqrt(2)/2}
\newcommand{\U}[3]{
 \draw[very thick, shift={(#2,#3)}, rotate=#1] (.9,.9)--(1.25,1.25);
 \draw[very thick, shift={(#2,#3)}, rotate=#1] (.1,.1)--(-.25,-.25);
 \draw[very thick, shift={(#2,#3)}, rotate=#1] (.9,.1)--(1.25,-.25);
 \draw[very thick, shift={(#2,#3)}, rotate=#1] (.1,.9)--(-.25,1.25);
 \fill[\darkred, rounded corners, shift={(#2,#3)}, rotate=#1] (0, 0) rectangle (1, 1) {};
 \draw[very thick, rounded corners, shift={(#2,#3)}, rotate=#1] (0, 0) rectangle (1, 1) {};
 \draw[thick,shift={(#2,#3)}, rotate=#1] (.65,.85)-- (.85,.85)-- (.85,.65);}

\newcommand{\hasse}{\draw[very thick] (1,0) -- (0,3) ; 
 \draw[very thick] (1,0) -- (2,3) ; 
 \draw[very thick] (1,0) -- (6,3) ; 
 \draw[very thick] (1,0) -- (4,3) ; 
 \draw[very thick] (1,0) -- (-2,3) ; 
 \draw[very thick] (1,0) -- (-4,3) ; 
 \draw[very thick] (1,9) -- (0,6) ; 
 \draw[very thick] (1,9) -- (2,6) ; 
 \draw[very thick] (1,9) -- (4,6) ; 
 \draw[very thick] (1,9) -- (6,6) ; 
 \draw[very thick] (1,9) -- (-2,6) ; 
 \draw[very thick] (1,9) -- (-4,6) ; 
 \draw[very thick] (-4,3) -- (-4,6) ; 
 \draw[very thick] (-4,3) -- (-2,6) ; 
 \draw[very thick] (-4,3) -- (0,6) ; 
 \draw[very thick] (-2,3) -- (-4,6) ; 
 \draw[very thick] (-2,3) -- (2,6) ; 
 \draw[very thick] (-2,3) -- (4,6) ; 
 \draw[very thick] (0,3) -- (-2,6) ; 
 \draw[very thick] (0,3) -- (4,6) ; 
 \draw[very thick] (2,3) -- (-4,6) ; 
 \draw[very thick] (2,3) -- (6,6) ; 
 \draw[very thick] (4,3) -- (6,6) ;
 \draw[very thick] (4,3) -- (2,6) ;
 \draw[very thick] (4,3) -- (-2,6) ;
 \draw[very thick] (6,3) -- (6,6) ;
 \draw[very thick] (6,3) -- (4,6) ;
 \draw[very thick] (6,3) -- (-2,6) ;
 \NCd{1}{0}
 \NCd{0}{3}
 \NCac{0}{3}
 \NCd{2}{3}
 \NCbd{2}{3}
 \NCd{4}{3}
 \NCab{4}{3}
 \NCd{-2}{3}
 \NCcd{-2}{3}
 \NCd{-4}{3}
 \NCbc{-4}{3}
 \NCd{6}{3}
 \NCad{6}{3}
 \NCd{0}{6}
 \NCad{0}{6}
 \NCbc{0}{6}
 \NCd{2}{6}
 \NCab{2}{6}
 \NCcd{2}{6}
 \NCd{4}{6}
 \NCac{4}{6}
 \NCcd{4}{6}
 \NCad{4}{6}
 \NCd{-2}{6}
 \NCac{-2}{6}
 \NCbc{-2}{6}
 \NCab{-2}{6}
 \NCd{-4}{6}
 \NCbc{-4}{6}
 \NCcd{-4}{6}
 \NCbd{-4}{6}
 \NCd{6}{6}
 \NCab{6}{6}
 \NCbd{6}{6}
 \NCad{6}{6}
 \NCd{1}{9}
 \NCab{1}{9}
 \NCbc{1}{9}
 \NCcd{1}{9}
 \NCad{1}{9}}
\newcommand{\Ud}[3]{
 \draw[very thick,shift={(#2,#3)}, rotate=#1] (.9,.9)--(1.25,1.25);
 \draw[very thick,shift={(#2,#3)}, rotate=#1] (.1,.1)--(-.25,-.25);
 \draw[very thick,shift={(#2,#3)}, rotate=#1] (.9,.1)--(1.25,-.25);
 \draw[very thick,shift={(#2,#3)}, rotate=#1] (.1,.9)--(-.25,1.25);
 \fill[\darkblue, rounded corners,shift={(#2,#3)}, rotate=#1] (0, 0) rectangle (1, 1) {};
 \draw[very thick, rounded corners,shift={(#2,#3)}, rotate=#1] (0, 0) rectangle (1, 1) {};
 \draw[thick,shift={(#2,#3)}, rotate=#1] (.65,.85)-- (.85,.85)-- (.85,.65);}

\newcommand{\dotchunk}[1]{\filldraw[\darkred] (0+#1,0) circle (3pt);
 \filldraw[\darkred] (.5+#1,0) circle (3pt);
 \node[] at (1+#1,0) {$\dots$};
 \filldraw[\darkred] (1.5+#1,0) circle (3pt);
 \filldraw[\darkred] (2+#1,0) circle (3pt);
 \filldraw[\darkblue] (2.5+#1,0) circle (3pt);
 \filldraw[\darkblue] (3+#1,0) circle (3pt);
 \node[] at (3.5+#1,0) {$\dots$};
 \filldraw[\darkblue] (4+#1,0) circle (3pt);
 \filldraw[\darkblue] (4.5+#1,0) circle (3pt);}

\newcommand{\wick}[3]{\draw[thick] (#1,0) -- (#1,{-.25*#3}) -- (#2, {-.25*#3}) -- (#2, 0); 
}

\newcommand{\wickflip}[5]{\draw[thick] ({#1+#4},#5) -- ({#1+#4},{.25*#3+#5}) -- ({#2+#4}, {.25*#3+#5}) -- ({#2+#4}, #5); 
}

\newcommand{\wickU}[3]{\draw[very thick] (#1,-1.5) -- (#1,{-.25*#3-1.5}) -- (#2, {-.25*#3-1.5}) -- (#2, 0-1.5); 
}

\newcommand{\wickflipU}[3]{\draw[very thick] (#1,0-0) -- (#1,{.25*#3-0}) -- (#2, {.25*#3-0}) -- (#2, 0-0); 
}

\newcommand{\NCd}[2]{
 \filldraw[\lightblue] (0+#1,0+#2) circle (20pt) ;
 \draw[very thick] (0+#1,0+#2) circle (20pt) ;
 \filldraw[] (.7+#1,0+#2) circle (3pt);
 \filldraw[] (0+#1,.7+#2) circle (3pt);
 \filldraw[] (-.7+#1,0+#2) circle (3pt);
 \filldraw[] (0+#1,-.7+#2) circle (3pt);}

\newcommand{\NCab}[2]{\draw[very thick] (-.7+#1,0+#2) -- (0+#1,-.7+#2) ;}
\newcommand{\NCac}[2]{\draw[very thick] (-.7+#1,0+#2) -- (.7+#1,0+#2); }
\newcommand{\NCad}[2]{\draw[very thick] (-.7+#1,0+#2)-- (0+#1,.7+#2) ;}
\newcommand{\NCbc}[2]{\draw[very thick] (0+#1,-.7+#2) -- (.7+#1,0+#2);}
\newcommand{\NCbd}[2]{\draw[very thick] (0+#1,-.7+#2) -- (0+#1,.7+#2); }
\newcommand{\NCcd}[2]{\draw[very thick] (.7+#1,0+#2) -- (0+#1,.7+#2); }

\newcommand{\tblock}[3]{ 
\U{-135+#1}{0+#2}{0+#3} \U{-135+#1}{4+#2}{0+#3} \U{-135+#1}{6+#2}{0+#3} 
\node[] at (2+#2,{\rt+#3}) {$\dots$};
\Ud{-45+#1}{{8+#2+\rt}}{{\rt+#3}}\Ud{-45+#1}{{12+\rt+#2}}{{\rt+#3}}\Ud{-45}{{14+\rt+#2}}{{\rt+#3}}
\node[] at (10+#2,{\rt+#3}) {$\dots$};}

\newcommand{\tblocknomid}[3]{ 
\U{-135+#1}{0+#2}{0+#3} \U{-135+#1}{4+#2}{0+#3} 
\node[] at (2+#2,{\rt+#3}) {$\dots$};
\Ud{-45+#1}{{12+\rt+#2}}{{\rt+#3}}\Ud{-45}{{14+\rt+#2}}{{\rt+#3}}
\draw[very thick] (5,{\rt+#3}) -- (9,{\rt+#3});
\node[] at (10+#2,{\rt+#3}) {$\dots$};
}

\newcommand{\tblocknoout}[3]{ 
\U{-135+#1}{4+#2}{0+#3} \U{-135+#1}{6+#2}{0+#3} 
\node[] at (2+#2,{\rt+#3}) {$\dots$};
\Ud{-45+#1}{{8+#2+\rt}}{{\rt+#3}}\Ud{-45+#1}{{12+\rt+#2}}{{\rt+#3}}
\draw[very thick] (0,{\rt+#3-1})-- (0,{\rt+#3}) -- (1,{\rt+#3});
\draw[very thick] (14,{\rt+#3-1})-- (14,{\rt+#3}) -- (13,{\rt+#3});
\node[] at (10+#2,{\rt+#3}) {$\dots$};}

\newcommand{\tblockshortflip}[3]{ 
\Ud{-45+#1}{{0+#2+\rt}}{{0+#3+\rt}} \Ud{-45+#1}{{4+#2+\rt}}{{0+#3+\rt}} 
\node[] at (2+#2,{\rt+#3}) {$\dots$};
\U{-135+#1}{{6+\rt+#2-\rt}}{{\rt+#3-\rt}}\U{-135}{{10+\rt+#2-\rt}}{{\rt+#3-\rt}}
\node[] at (8+#2,{\rt+#3}) {$\dots$};
}

\def\centerarc[#1](#2)(#3:#4:#5)
 { \draw[#1] ($(#2)+({#5*cos(#3)},{#5*sin(#3)})$) arc (#3:#4:#5); }

\usepackage{bbm}

\begin{document}
\count\footins = 1000

\title{Free Independence and the Noncrossing Partition Lattice in Dual-Unitary Quantum Circuits}

\author{Hyaline Junhe Chen}
\email{hc9760@princeton.edu}
\affiliation{Department of Physics, Princeton University, Princeton, NJ 08544}

\author{Jonah Kudler-Flam}%
 \email{jkudlerflam@ias.edu}
 \affiliation{School of Natural Sciences, Institute for Advanced Study, Princeton, NJ 08540, USA}
 \affiliation{Princeton Center for Theoretical Science, Princeton University, Princeton, NJ 08544, USA}

\begin{abstract}
We investigate details of the chaotic dynamics of dual-unitary quantum circuits by evaluating all $2k$-point out-of-time-ordered correlators. For the generic class of circuits, by writing the correlators as contractions of a class of quantum  channels, we prove their exponential decay. This implies that local operators separated in time approach \textit{free independence}. Along the way, we develop a replica trick for dual-unitary circuits, which may be useful and of interest in its own right. We classify the relevant eigenstates of the replica transfer matrix by paths in the lattice of noncrossing partitions, combinatorial objects central to free probability theory. Interestingly, the noncrossing lattice emerges even for systems without randomness and with small onsite Hilbert space dimension. 
\end{abstract}

\maketitle
\tableofcontents

\section{Introduction}

A fundamental difference between classical and quantum mechanical systems is that observables commute classically, while they generically do not commute quantum mechanically. In a quantum many-body system, spacelike separated observables, such as equal time Pauli operators, commute
\begin{align}
 [\sigma_{\alpha}(x,t), \sigma_{\gamma}(y,t)] = 0,\quad \alpha,\gamma \in \{x,y,z\}.
\end{align}
However, once they become in causal contact, their commutator grows
\begin{align}
 [\sigma_{\alpha}(x,0), \sigma_{\gamma}(y,t)] = O(1).
\end{align}
The transition in this behavior 
is a signature of quantum thermalization and quantum chaos. The thermal expectation value of the nontrivial piece of the square of this commutator  is the so-called out-of-time-ordered correlator (OTOC)
\begin{align}
 C^{(2)}(x,y,t) \defn \langle \sigma_{
 \alpha
 }(x,0) \sigma_{\gamma}(y,t) \sigma_{\alpha}(x,0) \sigma_{\gamma}(y,t)\rangle,
\end{align}
which has played a central role in modern understanding of quantum many-body chaos. The OTOC has given insights into diverse systems, such as quantum circuits \cite{2018PhRvX...8c1058R,2018PhRvX...8b1013V,2018PhRvX...8b1014N}, black holes \cite{2014JHEP...03..067S,2015NuPhB.901..382J, 2015PhRvL.115m1603R,2014JHEP...12..046S,2016JHEP...08..106M,2016PhRvD..94j6002M,2015JHEP...03..051R}, localizing spin systems \cite{2017AnP...52900318H, 2017SciBu..62..707F,2017PhRvB..95f0201S,2016arXiv160802765C}, and even quantum experiments \cite{2016PhRvA..94d0302S, 2016arXiv160701801Y, 2016PhRvA..94f2329Z, 2017PhRvE..95f2127C, 2017PhRvA..95a2120Y, 2024arXiv240501642L,2017NatPh..13..781G,2017PhRvX...7c1011L, 2016arXiv161205249W,2017arXiv170506714M}. At early times, the OTOC will equal unity because the Pauli operators commute and square to the identity operator. At late times, if the dynamics are sufficiently chaotic, the OTOC will decay to zero due to the eigenvectors of the two operators being sufficient randomized relative to one another.
While the four-point OTOC, $C^{(2)}$, provides significant information, it is far from the entire story\footnote{Indeed, there are known cases where the OTOC spuriously appears chaotic \cite{2020PhRvL.124n0602X, 2020PhRvE.101a0202P,2018PhRvB..98m4303P,2018arXiv181209237H,2021ScPC....4....2B,2021ScPC....4....2B,2020JHEP...11..068H}.}; this is analogous to the fact that merely knowing the variance of a probability distribution does not fully characterize it.
If the dynamics are truly chaotic, then dynamics will cause the eigenvectors of the two operators to be randomized and \textit{all} k-point OTOCs will decay
\begin{align}
\label{eq:deeptherm}
 C^{(k)}(x,y,t) \defn \langle \left( \sigma_{\alpha}(x,0)\sigma_{\gamma}(y,t)\right)^k\rangle \rightarrow 0.
\end{align}
High point correlation functions are generally quite difficult to evaluate, so this strong form of chaos, while expected to be generic, has not to our knowledge been explicitly verified in any physical quantum systems. One of our main results is proving exponential decay of all $2k$-point OTOCs in a large class of analytically tractable quantum circuit models, namely the generic subclass of so-called dual-unitary circuits \cite{2018PhRvL.121z4101B, 2019PhRvL.123u0601B, 2019PhRvB.100f4309G}. This may be considered a strong form of quantum information scrambling and thermalization, which has recently been the topic of significant research in the context of so-called \textit{deep thermalization} \cite{2023PRXQ....4a0311C, 2023Natur.613..468C,2022PhRvL.128f0601H,2022Quant...6..738C,2022Quant...6..886I,2024arXiv240311970M}.

Interestingly, this strong form of thermalization has been studied in parallel in the mathematical literature under the guise of free probability theory \cite{nica2006lectures, mingo2017free}. Free probability theory is the generalization of classical probability theory to noncommutative operators, so it is natural that it is the proper language for chaotic quantum systems. This perspective has recently been advocated for in the context of the eigenstate thermalization hypothesis \cite{2022PhRvL.129q0603P, 2023arXiv230806200F, 2023arXiv230300713P,2024arXiv240901404F}.\footnote{Free probability theory has also played an important role in recent developments in tensor networks \cite{2010JPhA...43A5303C,2021PRXQ....2d0340K,2022JHEP...02..076K,2024AnHP...25.2107C}, disordered condensed matter systems \cite{2012PhRvL.109c6403C,2010arXiv1012.5039M,2011PhRvL.107i7205M,2023PhRvX..13a1045H}, quantum information theory \cite{2016JMP....57a5215C}, and quantum black holes \cite{2019arXiv191111977P,2023JHEP...10..040W}.} Given a state $\varphi$, which is a linear functional on an algebra of operators, two algebra elements $a$ and $b$ are \textit{freely independent} if 
\begin{align}
\label{eq:freeness}
 \varphi(f_1(a) g_1(b)f_2(a) g_2(b)\dots f_n(a) g_n(b)) = 0, \quad \forall n
\end{align}
whenever $\varphi(f_i(a)) = \varphi(g_i(b)) = 0$, where $f_i$ and $g_i$ are polynomials of the operators. Upon identifying $a$ and $b$ with the above Pauli operators and $\varphi$ with the thermal expectation value, this can be seen to be equivalent to \eqref{eq:deeptherm}.
Free independence may be contrasted with \textit{tensor independence} where
\begin{align}
 \varphi(f_1(a) g_1(b)f_2(a) g_2(b)\dots f_n(a) g_n(b)) = \varphi(f_1(a)f_2(a)\dots f_n(a) )\times \varphi( g_1(b)g_2(b)\dots g_n(b)).
\end{align}
Tensor independence is what occurs for operators that are at spacelike separation. Therefore, we may rephrase the notion of quantum thermalization as the transition from tensor independence to free independence.

Free independence of operators has been proven in a variety of cases of independent random matrices as the size of the matrix becomes large \cite{mingo2017free}. While of clear mathematical interest, these examples do not directly correspond to the notion of chaos that we explore because the emergence of free independence is not from quantum dynamics. Interestingly, it has been proven that deterministic variables become asymptotically free if they are evolved by a random Wigner matrix \cite{CIPOLLONI2022109394}\footnote{Wigner matrices are Hermitian matrices with independent and identically distributed matrix elements and the OTOCs were studied in further detail in \cite{2024arXiv240217609C}. The specification to Gaussian random matrices can be gleaned from earlier conjectures in \cite{Cotler_2017}.}. This was an important first step in understanding how chaotic dynamics lead to freeness. The improvement in the current work is that we work with local quantum systems and without disorder, which is the setting of physical quantum systems.

In the remainder of the introduction, we review the structure of dual-unitary circuits, demonstrating how certain calculations become analytically tractable. We then review the combinatorics of noncrossing partitions, their associated lattice, and their relation to free probability theory. In Section \ref{sec:eig}, we develop a replica trick for dual-unitary circuits by classifying the eigenstates of the replica transfer matrices using paths in the noncrossing partition lattice. This technical contribution may be applied to many different calculations, not necessarily related to free independence. In Section \ref{sec:otoc}, we use the results of the prior section to explicitly evaluate the OTOCs for dual-unitary circuits, proving that they can be expressed in terms of a class of quantum channels, and consequently exponentially decay at late times. There, we specify to systems with large onsite Hilbert space dimension to simplify calculations. In Section \ref{sec:approacj}, we numerically study the self-dual kicked Ising model, which has onsite Hilbert space dimension $2$ and can be integrable or chaotic depending on the model parameters. We observe the generic approach to free independence by evaluating the spectra of sums of Pauli operators. The spectra approach the so-called free convolution of Bernoulli distributions. At the integrable point, the spectra remain as the classical convolution of Bernoulli distributions for all times. In Section \ref{sec:disc}, we comment on several open questions.

\subsection{Dual-Unitary Circuits}

\begin{figure}
 \begin{tikzpicture}
 \U{0}{0}{0} \U{0}{1.5}{1.5} 
 \U{0}{3}{0} \U{0}{4.5}{1.5} 
 \U{0}{6}{0} \U{0}{7.5}{1.5} 
 \U{0}{9}{0} \U{0}{10.5}{1.5} 
 \U{0}{0}{3} \U{0}{1.5}{4.5} 
 \U{0}{3}{3} \U{0}{4.5}{4.5} 
 \U{0}{6}{3} \U{0}{7.5}{4.5} 
 \U{0}{9}{3} \U{0}{10.5}{4.5} 
 \node[] at (-1,3) {$\dots$};
 \node[] at (12.5,3) {$\dots$};
 \end{tikzpicture}
 \caption{A brickwork circuit of two-site unitaries.}
 \label{fig:brickwork}
\end{figure}
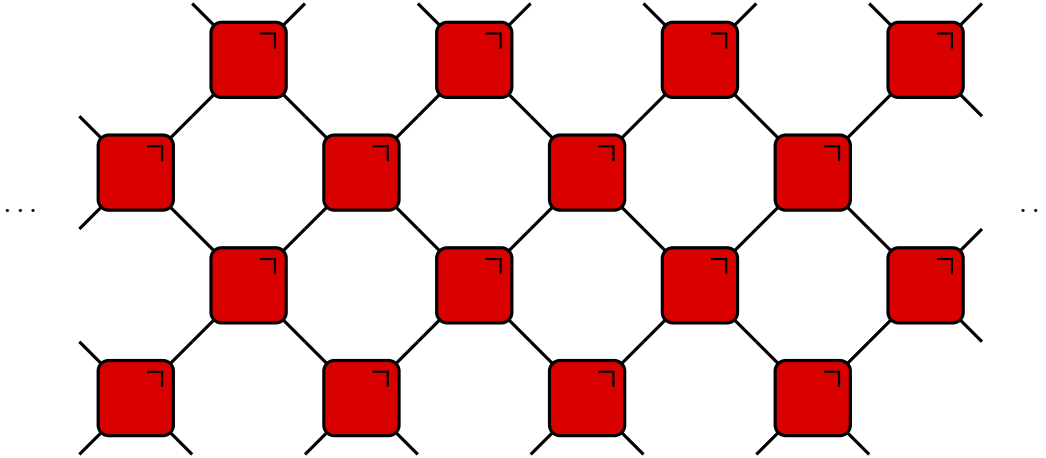

We consider quantum circuits composed of unitary operators acting on two sites of local Hilbert space dimension $q$, i.e.~unitary operators on $\mathbb{C}_q \otimes \mathbb{C}_q$. For simplicity, we take $q$ to be a power of $2$ such that generalized Pauli operators, consisting of the tensor products of $2\times 2$ Pauli operates, form a natural basis. The diagrammatic representation of the circuit is extremely useful in visualization and computation. We represent the unitary, $U$, as
\begin{align}
 \begin{tikzpicture}
 \U{0}{0}{0}
 \end{tikzpicture}
\end{align}
and its adjoint as 
\begin{align}
 \begin{tikzpicture}
 \Ud{0}{0}{0}
 \end{tikzpicture}.
\end{align}
The unitarity of these operators is expressed diagrammatically as
\begin{align}
 \begin{tikzpicture}
 \Ud{0}{-5}{0} \U{0}{-5}{-2}
 \draw[very thick] (-5.25,-.25) -- (-5.25,-.75) ;
 \draw[very thick] (1.25-5,-.25) -- (1.25-5,-.75) ;
 \node[scale = 2] at (-2,-.5) {$=$}; 
 \U{0}{0}{0} \Ud{0}{0}{-2}
 \draw[very thick] (-.25,-.25) -- (-.25,-.75) ;
 \draw[very thick] (1.25,-.25) -- (1.25,-.75) ;
 \node[scale = 2] at (3,-.5) {$=$}; 
 \draw[very thick] (6,-2.25) -- (6,1.25) ;
 \draw[very thick] (5,-2.25) -- (5,1.25) ;
 \end{tikzpicture}
 \label{eq:unit}
\end{align}
where the solid lines are identity operators.
To model quantum many-body dynamics, the unitaries are concatenated into a global unitary via a circuit with a brickwork geometry, as sketched in Figure \ref{fig:brickwork}. The total system has $L$ sites, and we will often take $L = \infty$. 

We consider a subset of unitary circuits where the operators obey a ``dual-unitary'' relation, which is diagrammatically represented as 
\begin{align}
 \begin{tikzpicture}
 \U{0}{0}{0} \Ud{0}{0}{-2}
 \draw[very thick] (-.25,-.25) -- (-.25,-.75) ;
 \draw[very thick] (-.25,1.25)--(-.5,1.25) -- (-.5,-2.25)--(-.25,-2.25) ;
 \node[scale = 2] at (3,-.5) {$=$}; 
 \draw[very thick] (6,1.25) -- (5.75,1) -- (4.75,1)-- (4.75,-2.25)-- (5.75,-2.25)-- (6,-2.5);
 \draw[very thick] (6,-.25)-- (5.75,0)-- (5.25,0)-- (5.25,-1.25) --(5.75,-1.25)--(6,-1) ;
 \U{0}{8}{0} \Ud{0}{8}{-2}
 \draw[very thick] (-.25+9.5,-.25) -- (-.25+9.5,-.75) ;
 \draw[very thick] (-.25+9.5,1.25)--(-.5+10,1.25) -- (-.5+10,-2.25)--(-.25+9.5,-2.25) ;
 \node[scale = 2] at (11,-.5) {$=$}; 
 \draw[very thick] (4.5+8,1.25) -- (4.75+8,1) -- (5.75+8,1)-- (5.75+8,-2.25)-- (4.75+8,-2.25)-- (4.5+8,-2.5);
 \draw[very thick] (5.25+7.25,-.25)-- (5.5+7.25,0)-- (6+7.25,0)-- (6+7.25,-1.25) --(5.5+7.25,-1.25)--(5.25+7.25,-1) ;
 \end{tikzpicture}
 \label{eq:dualunit}.
\end{align}

Dual unitarity adds enough structure to the models such that many questions are answerable analytically. However, they are not so restrictive as to trivialize dynamics, allowing for a wide range of dynamical behavior, ranging from integrable to fully chaotic \cite{2019PhRvL.123u0601B}.
The non-generic aspect of dual-unitary circuits is that infinite temperature two-point correlation functions are only non-vanishing on the light cone
\begin{align}
 \langle \sigma_{\alpha}(0,0) \sigma_{\gamma}(x,t) \rangle = \delta_{x,t} f_{\alpha \gamma}(t),
\end{align}
where $f(t)$ is a yet to be determined function. 
This can be seen directly by repeatedly using the unitarity and dual unitarity relations. Off the light cone, the circuit contracts such that it is proportional to the trace of the Pauli operator, i.e.~zero. On the light cone, 
\begin{align}
\scalebox{.5}{
 \begin{tikzpicture}
 \U{0}{0}{0} \U{0}{1.5}{1.5} 
 \U{0}{3}{0} \U{0}{4.5}{1.5} 
 \U{0}{6}{0} \U{0}{7.5}{1.5} 
 \U{0}{9}{0} \U{0}{10.5}{1.5} 
 \U{0}{0}{3} \U{0}{1.5}{4.5} 
 \U{0}{3}{3} \U{0}{4.5}{4.5} 
 \U{0}{6}{3} \U{0}{7.5}{4.5} 
 \U{0}{9}{3} \U{0}{10.5}{4.5} 
 \node[] at (-1,3-3) {$\dots$};
 \node[] at (12.5,3-3) {$\dots$};
 \Ud{0}{1.5}{-6} \Ud{0}{3}{-4.5} 
 \Ud{0}{4.5}{-6} \Ud{0}{6}{-4.5} 
 \Ud{0}{7.5}{-6} \Ud{0}{9}{-4.5} 
 \Ud{0}{10.5}{-6} \Ud{0}{0}{-4.5} 
 \Ud{0}{1.5}{-3} \Ud{0}{3}{-1.5} 
 \Ud{0}{4.5}{-3} \Ud{0}{6}{-1.5} 
 \Ud{0}{7.5}{-3} \Ud{0}{9}{-1.5} 
 \Ud{0}{10.5}{-3} \Ud{0}{0}{-1.5} 
 \filldraw[green] (10.25,-.25) circle (5pt);
 \draw[very thick] (10.25,-.25) circle (5pt);
 \filldraw[green] (4.25,-6.25) circle (5pt);
 \draw[very thick] (4.25,-6.25) circle (5pt);
 \node[scale = 3] at (15,0) {$\Rightarrow$};
 \U{0}{7.5+14}{1.5} 
 \U{0}{9+14}{0} 
 \U{0}{4.5+14}{4.5} 
 \U{0}{6+14}{3} 
 \Ud{0}{4.5+14}{-6} \Ud{0}{6+14}{-4.5} 
 \Ud{0}{7.5+14}{-3} \Ud{0}{9+14}{-1.5} 
 \draw[very thick] (6+14-.25,3-.25) -- (6+14-.25,-3-.25);
 \draw[very thick] (7.5+14-.25,3-1.75) -- (7.5+14-.25,-3+1.25);
 \draw[very thick] (6+14-1.75,3-.25+1.5) -- (6+14-1.75,-4.75);
 \draw[very thick] (6+14-1.75,3-.25+3) --(6-1+14-1.75,3-.25+3) -- (6-1+14-1.75,-4.75-1.5)-- (6+14-1.75,-4.75-1.5);
 \draw[very thick] (10.25+14, 1.25) -- (10.25+14+1, 1.25) -- (10.25+14+1, -1.75) -- (10.25+14, -1.75) ;
 \draw[very thick] (10.25+14-1.5, 2.75) -- (10.25+14+2, 2.75) -- (10.25+14+2, -1.75-1.5) -- (10.25+14-1.5, -1.75-1.5) ;
 \draw[very thick] (10.25+14-1.5-1.5, 1.25+3) -- (10.25+14+3, 1.25+3) -- (10.25+14+3, -1.75-1.5-1.5) -- (10.25+14-1.5-1.5, -1.75-1.5-1.5) ;
 \draw[very thick] (10.25+14-1.5-1.5-1.5, 1.25+4.5) -- (10.25+14+4, 1.25+4.5) -- (10.25+14+4, -1.75-1.5-1.5-1.5) -- (10.25+14-1.5-1.5-1.5, -1.75-1.5-1.5-1.5) ;
 \filldraw[green] (10.25+14,-.25) circle (5pt);
 \draw[very thick] (10.25+14,-.25) circle (5pt);
 \filldraw[green] (4.25+14,-6.25) circle (5pt);
 \draw[very thick] (4.25+14,-6.25) circle (5pt);
 \end{tikzpicture}}.
\end{align}
The trace is implemented by connecting the free tensor indices on the top and bottom of the diagram. The ``$\dots$'' represents the fact that we can be working with a system of arbitrarily large spatial extension.
The two-point function reduces to 
\begin{align}
 f_{\alpha \gamma}(t) =\frac{1}{q} \Tr\left( \sigma_{\gamma} \mathcal{M}^t_-(\sigma_{\alpha})\right)=\frac{1}{q} \Tr\left( \sigma_{\alpha} \mathcal{M}^t_+(\sigma_{\gamma})\right),
\end{align}
where $\mathcal{M}_+$ and $\mathcal{M}_-$ are defined as
\begin{align}
 \mathcal{M}_+(\sigma) = \Tr_1 \left( U (\sigma \otimes \mathbbm{1})U^{\dagger}\right),\quad\mathcal{M}_-(\sigma) = \Tr_2 \left( U^{\dagger} ( \mathbbm{1}\otimes \sigma)U\right).
\end{align}
The subscripts on the trace denote that these are partial traces over the first and second tensor factors respectively.
In diagrammatic notation 
 \begin{align}
 \label{eq:channel1}
 \begin{aligned}
\mathcal{M}_{-}(\sigma) \equiv \frac{1}{q} \quad
\scalebox{.7}{
 \begin{tikzpicture}[baseline=-3ex]
 \U{0}{0}{0} \Ud{0}{0}{-2}
 \draw[very thick] (-.25,-.25) -- (-.25,-.75) ;
 \draw[very thick] (1.25,-.25) -- (1.25,-.75) ;
 \filldraw[green] (1.25,-.5) circle (5pt);
 \draw[very thick] (1.25,-.5) circle (5pt);
 \draw[very thick] (1.25,1.25) -- (1.75,1.25) -- (1.75,-2.25) -- (1.25,-2.25);
 \node[scale = 1] at (.7,-0.5) { $\sigma$};
 \end{tikzpicture}}
 \quad = \quad \frac{1}{q} \quad
 \scalebox{.7}{
 \begin{tikzpicture}[baseline=-3ex]
 \Ud{-45}{{\rt}}{{\rt}}\U{-135}{{2}}{{0}}
 \draw[very thick] (0,-1.5) -- (0,-2)--(2,-2) --(2,-1.5);
 \draw[very thick] (0,{1+\rt}) -- (0,{1.5+\rt})--(2,{1.5+\rt}) --(2,{1+\rt});
 \filldraw[green] (1,{\rt}) circle (5pt);
 \draw[very thick] (1,{\rt}) circle (5pt);
 \node[scale = 1] at (1,-.2) { $\sigma$};
 \end{tikzpicture}}\\
 \mathcal{M}_{+}(\sigma) \equiv \frac{1}{q} \quad
 \scalebox{.7}{
 \begin{tikzpicture}[baseline=-3ex]
 \Ud{0}{0}{0} \U{0}{0}{-2}
 \draw[very thick] (-.25,-.25) -- (-.25,-.75) ;
 \draw[very thick] (1.25,-.25) -- (1.25,-.75) ;
 \filldraw[green] (-.25,-.5) circle (5pt);
 \draw[very thick] (-.25,-.5) circle (5pt);
 \draw[very thick] (-.25,1.25) -- (-.75,1.25) -- (-.75,-2.25) -- (-.25,-2.25);
 \node[scale = 1] at (0.3,-0.5) { $\sigma$};
 \end{tikzpicture}}
 \quad = \quad \frac{1}{q} \quad
 \scalebox{.7}{
 \begin{tikzpicture}[baseline=-3ex]
 \U{-135}{0}{0}\Ud{-45}{{2+\rt}}{{\rt}}
 \draw[very thick] (0,-1.5) -- (0,-2)--(2,-2) --(2,-1.5);
 \filldraw[green] (1,{\rt}) circle (5pt);
 \draw[very thick] (1,{\rt}) circle (5pt);
 \node[scale = 1] at (1,-.2) { $\sigma$};
 \draw[very thick] (0,{1+\rt}) -- (0,{1.5+\rt})--(2,{1.5+\rt}) --(2,{1+\rt});
 \end{tikzpicture}}
 \end{aligned}
\end{align}
These maps are quantum channels i.e.~completely positive and trace preserving (CPTP). Moreover, they are unital, mapping the identity operator to itself. These are known as \textit{unistochastic maps} \cite{2004quant.ph..1119Z,2013PhRvA..87b2111M}. Such maps have eigenvalues that lie on the unit disc, and the classification of the ergodicity of dual-unitary circuits can thus be made by the associated spectra \cite{2019PhRvL.123u0601B}. In this paper, we will mainly be considered with the generic setting where the dynamics are both ergodic and mixing, with all eigenvalues lying within the disc. There is just one eigenvalue lying on the unit circle, corresponding to the identity operator. Because $\sigma_{\alpha}$ and $\sigma_{\gamma}$ have no support on the identity operator, $\Tr(\sigma_\alpha \mathbbm{1})=\Tr(\sigma_\gamma \mathbbm{1}) = 0$, the unistochasticity leads to an exponential decay (in $t$) of the two-point function. When considering higher-point correlations functions, we will encounter generalized unistochastic maps which enable us to similarly prove the exponential decay of all higher-point correlation functions. These higher-point correlation functions will depend on the structure of replica transfer matrices, which we now define.

The unreplicated transfer matrix is composed of $n$ unitary operators and $n$ dual-unitary operators
\begin{equation}
 T^{(1)}_n = \frac{1}{q}
 \times 
 \scalebox{.7}{
 \begin{tikzpicture}[baseline=-3.5ex]
 \tblock{0}{0}{0};
 \draw[very thick] ({-1.03},{\rt}) -- (-1.03,-3) -- (15.03,-3) -- (15.03,{\rt}) ;
 \draw[thick ,decoration={
 brace,
 raise=-0.5cm
 },decorate](-1,1) -- (6.5,1) 
node [pos=0.5,yshift=0cm, scale = 1.5] {$n$}; 
 \draw[thick ,decoration={
 brace,
 raise=-0.5cm
 },decorate](7.5,1) -- (15,1) 
node [pos=0.5,yshift=0cm,scale = 1.5] {$n$}; 
 \end{tikzpicture}}.
\end{equation}
It is important to note the orientation of each operator. The replica transfer matrix is constructed by horizontally concatenating the transfer matrix $k$ times as
\begin{equation}
\label{eq:tkn}
T^{(k)}_n =\frac{1}{q}
 \times 
\scalebox{0.2}{\begin{tikzpicture}[baseline=-6ex]
 \tblockshortflip{0}{0}{0};
 \tblockshortflip{0}{12}{0};
 \tblockshortflip{0}{24+18}{0};
 \tblockshortflip{0}{36+18}{0};
 \draw[very thick] (-1,{\rt}) -- (-1,{-.25*11-1.5}) -- (10+36+18+1, {-.25*11-1.5}) -- (10+36+18+1, {\rt});
 \node[scale = 3] at (32,-0.5) { $\textbf{\dots}$};
 \draw [very thick, decorate,decoration={brace,amplitude=6pt,mirror,raise=4ex}]
 (10,0) -- (0,0) node[midway,yshift=4em,scale=4]{$k$ of these};
 \draw [decorate,decoration={brace,amplitude=5pt,mirror,raise=11ex}]
 (0,0) -- (4,0) node[midway,yshift=-7em,scale=4]{$n$};
 \end{tikzpicture}}.
\end{equation}
Section \ref{sec:eig} will be devoted to determining the structure of the eigenvectors of this operator. 

These transfer matrices, for $k=1$ and $2$, have previously shown up in the computations of $4$-point OTOCs \cite{2020PhRvR...2c3032C} and in the calculation of operator entanglement \cite{2020ScPP....8...67B}. We will show that the generalized higher $k$ transfer matrices appear in the computation of higher point OTOCs. 

\subsection{The Noncrossing Partition Lattice}

Consider the set of partitions of the numbers $1\dots k \equiv [k]$. There is a partial order on the set given by refinement. Namely, $\tau_1 \leq \tau_2$ if the ``blocks'' of $\tau_1$ are contained within the blocks of $\tau_2$. For example, for $k = 4$
\begin{align}
 \underbrace{[1,3] [2] [4]}_{\tau_1} \leq \underbrace{[1,3,4] [2]}_{\tau_2},
\end{align}
where we use the canonical form of the partitions where the smallest numbers are listed first in each block. 

A partition is called noncrossing if whenever $a<b<c<d \in [k]$, $a$ and $c$ are in the same block, and $b$ and $d$ are in the same block, then all four elements are in the same block. Arranging the elements on a circle and connecting the elements of each block, the noncrossing partitions are seen to have no crossings. For example, the partition $[1,2] [3,4]$
\begin{align}
 \begin{tikzpicture}
 \draw[very thick] (0,0) circle (20pt) ;
 \filldraw[] (.7,0) circle (3pt);
 \filldraw[] (0,.7) circle (3pt);
 \filldraw[] (-.7,0) circle (3pt);
 \filldraw[] (0,-.7) circle (3pt);
 \draw[very thick] (.7,0) -- (0,.7);
 \draw[very thick] (-.7,0) -- (0,-.7);
 \node[] at (-1,0) {$1$};
 \node[] at (0,-1) {$2$};
 \node[] at (1,0) {$3$};
 \node[] at (0,1) {$4$};
 \end{tikzpicture}
\end{align}
is noncrossing, while $[1,3] [2,4]$
\begin{align}
 \begin{tikzpicture}
 \draw[very thick] (0,0) circle (20pt) ;
 \filldraw[] (.7,0) circle (3pt);
 \filldraw[] (0,.7) circle (3pt);
 \filldraw[] (-.7,0) circle (3pt);
 \filldraw[] (0,-.7) circle (3pt);
 \draw[very thick] (.7,0) -- (-.7,0);
 \draw[very thick] (0,-.7) -- (0,.7);
 \node[] at (-1,0) {$1$};
 \node[] at (0,-1) {$2$};
 \node[] at (1,0) {$3$};
 \node[] at (0,1) {$4$};
 \end{tikzpicture}
\end{align}
is crossing. The set of noncrossing partitions also admits a partial order via refinement. This partially ordered set has a lattice structure, meaning that every pair of partitions has a unique least upper bound and greatest lower bound. There exists both a minimal element $[1][2]\dots[k]$ and a maximal element $[1,2,\dots ,k]$ of the set. Moreover, the lattice is endowed with a rank function, $r$, such that if $\tau_1 < \tau_2$, then $r(\tau_1) < r(\tau_2)$. This is given by $k$ minus the number of blocks in the partition. For example, $r([1,3][2][4]) = 4-3 =1$. The lattice can be represented via a Hasse diagram whose vertices are elements of the set and edges connect comparable elements. The Hasse diagram is vertically oriented according to the rank function. For example, the Hasse diagram for $NC_4$ is
\begin{align}
\scalebox{.9}{
 \begin{tikzpicture}
 \hasse
 \node[] at (-8,0) {$r = 0$};
 \node[] at (-8,3) {$r = 1$};
 \node[] at (-8,6) {$r = 2$};
 \node[] at (-8,9) {$r = 3$};
 \end{tikzpicture}}
\end{align}
Within the lattice, one may consider chains of noncrossing partitions, that is, paths in the noncrossing lattice $\tau_1 \leq \tau_2 \leq \dots \leq \tau_k$. For example, $[1][2][3][4]\leq [1,2][3][4]\leq [1,2,4] [3]\leq [1,2,3,4]$
\begin{align}
\scalebox{.9}{
 \begin{tikzpicture}
 \hasse
 \draw[line width=5pt, red] (1,0) -- (4,3) -- (6,6) -- (1,9);
 \end{tikzpicture}}
\end{align}
Significant work has been done on combinatorially classifying $n$-chains. For example,
the total number of $n$-chains in $NC_k$ is given by the Fuss-Catalan numbers \cite{2006math.....11106A}
\begin{align}
\label{eq:FC}
 FC_k^{(n)} = \frac{1}{kn +1}\binom{nk + k}{k},
\end{align}
a one parameter generalization of the famed Catalan numbers ($n=1)$.

The rank is a rather coarse grained characterization of a partition. A finer grained characterization is the number of noncrossing partitions of $[k]$ with partition structure $(1^{m_1}2^{m_2}\dots k^{m_k})$\footnote{This notation means that the partition has $m_i$ blocks of size $i$.}, which is given by
\begin{align}
\label{eq:Kre}
 \# NC_k(1^{m_1}2^{m_2}\dots k^{m_k}) = \frac{k!}{(k+1-\sum_i m_i)!\prod_i m_i!}.
\end{align} 
This can be further refined by introducing the ``Kreweras complement,'' $K(\pi)$, of a partition. This is the largest element of $NC_k$ such that, when interlacing the set $[k]$ with itself, $\pi \cup K(\pi)$ remains noncrossing. For example, if $ \pi = [1,2] [3,4]$, then $K(\pi) = [1][2,4][3]$
\begin{align}
 \begin{tikzpicture}
 \draw[very thick] (0,0) circle (20pt) ;
 \filldraw[] (.7,0) circle (3pt);
 \filldraw[] (0,.7) circle (3pt);
 \filldraw[] (-.7,0) circle (3pt);
 \filldraw[] (0,-.7) circle (3pt);
 \filldraw[red] ({.7*\rt},{.7*\rt}) circle (3pt);
 \filldraw[red] ({-.7*\rt},{.7*\rt}) circle (3pt);
 \filldraw[red] ({.7*\rt},{-.7*\rt}) circle (3pt);
 \filldraw[red] ({-.7*\rt},{-.7*\rt}) circle (3pt);
 \draw[very thick] (.7,0) -- (0,.7);
 \draw[very thick] (-.7,0) -- (0,-.7);
 \draw[very thick,red] ({.7*\rt},{-.7*\rt}) -- ({-.7*\rt},{.7*\rt});
 \node[] at (-1,0) {$1$};
 \node[] at (0,-1) {$2$};
 \node[] at (1,0) {$3$};
 \node[] at (0,1) {$4$};
 \node[red] at ({\rt},{\rt}) {$1$};
 \node[red] at ({-\rt},{\rt}) {$2$};
 \node[red] at ({\rt},{-\rt}) {$4$};
 \node[red] at ({-\rt},{-\rt}) {$3$};
 \end{tikzpicture}
\end{align} 
In general,
\begin{align}
 r(\pi) + r(K(\pi)) = k-1.
\end{align}
Therefore, the Kreweras complement provides a bijection between rows in the Hasse diagram.
The number of noncrossing partitions $\pi$ of type $(1^{r_1}2^{r_2}\dots k^{r_k})$ with $K(\pi)$ of type $(1^{q_1}2^{q_2}\dots k^{q_k})$ is
\begin{align}
 k \frac{(r_1 + \dots +r_k-1)!(q_1+\dots +q_k-1)!}{\prod_i r_i!q_i!}.
\end{align}

Finally, it is crucial to note that the set of noncrossing partitions can be identified with a subset of the permutation group $S_k$, which we call the noncrossing permutations. There is a metric on $S_k$ given by the minimal number of transpositions that it takes to go from one permutation to the other. The set of noncrossing permutations are defined as the set of permutations that lie on the geodesic between the identity permutation and the $\mathbb{Z}_k$ cyclic permutation. In cycle notation, these two permutations are given by
\begin{align}
 \mathbbm{1} = (1)(2)\dots (k) , \quad \mathbb{Z}_k = (1,2,\dots k).
\end{align}
There is an isomorphism between the noncrossing partitions and the noncrossing permutations by promoting the blocks of the noncrossing partitions to cycles of noncrossing permutations and vice versa. In the remainder, the phrasing of permutations will be more natural, though we will still refer to the noncrossing partition lattice.

\subsection{Free Probability Theory}

We have already defined the notion of free independence in \eqref{eq:freeness}. That $a$ and $b$ are freely independent is an incredibly strong notion, and fixes all joint moments from the moments of the individual elements. With the knowledge that two variables, $a$ and $b$, are freely independent, we review what more can be said about them.

We first recall the relation between moments and classical cumulants. For random variables $X_i$, the mixed moments are related to the cumulants as
\begin{align}
 \mathbb{E}(X_1 \dots X_n) = \sum_{\pi \in \Pi_n} \prod_{B \in \pi}\kappa({X_i: i\in |B|})
\end{align}
where the sum runs over all set partitions of the numbers $[n]$, $\Pi_n$, and the product runs over all blocks of each partition. This expression can be inverted as
\begin{align}
 \kappa(X_1, \dots , X_n) = \sum_{\pi \in \Pi_n}(|\pi|-1)!(-1)^{|\pi|-1}\prod_{B\in \pi}\mathbb{E}\left(\prod_{i\in B}X_i\right).
\end{align}
If two or more of the variables in the cumulant are tensor independent, then the cumulant vanishes.

The moment-cumulant relation can be deformed such that the sum is only over noncrossing partitions
\begin{align}
\label{eq:mom_cum}
 \varphi(X_1 \dots X_n) = \sum_{\pi \in NC_n} \prod_{B \in \pi}\tilde{\kappa}({X_i: i\in |B|}).
\end{align}
These deformed cumulants are called \textit{free cumulants} \cite{speicher1994multiplicative} and vanish whenever two or more variables are freely independent. 
The moment-cumulant relation may be inverted as
\begin{align}
 \kappa_n(X_1, \dots , X_n) = \sum_{\pi \in NC_n}\varphi_{\pi}\left(\prod_{i\in B}X_i\right)\mu(\pi,\mathbbm{1}_n ),
\end{align}
where $\mu$ is the Mobius function of $NC_n$ whose explicit form can be found in e.g.~\cite{nica2006lectures}.
The free cumulants play a central role in free probability theory, in particular in \textit{free harmonic analysis}, where the knowledge of the probability measures, $\mu_a$ and $\mu_b$, of freely independent variables, $a$ and $b$, fixes the measure of their sum $\mu_{a+b} = \mu_a\boxplus \mu_b$, where $\boxplus$ is the free additive convolution operation.

To implement this, we first introduce the \textit{Cauchy transform} of a measure
\begin{align}
 G(z) = \int_\mathbb{R} (z-t)^{-1} d\mu(t).
\end{align}
This fixes the so-called \textit{$\mathcal{R}$-transform}
\begin{align}
 \mathcal{R}(z) = \sum_{n = 1}^{\infty} \kappa_n(X_1, \dots , X_n) z^n
\end{align}
via the moment-cumulant relation
\begin{align}
 z = G(\mathcal{R}(z)+1/z).
\end{align}
The utility of the $\mathcal{R}$-transform is that they are additive under free convolution, due to their construction from free cumulants, namely
\begin{align}
 \mathcal{R}_{a+b} = \mathcal{R}_a + \mathcal{R}_b.
\end{align}
Therefore, knowing the spectra of two variables, the spectra of their sum can be evaluated by adding their $\mathcal{R}$-transforms, then transforming back. Similarly one can compute the spectra of the product of freely independent variables using a related analytic function called the $\mathcal{S}$-transform, though this will not play a role in the remainder of the paper.

\section{Eigenstates of the Transfer Matrix}

\label{sec:eig}

In this section, we derive the eigenstates of the transfer matrix with unit eigenvalue that are common to all dual-unitary models. The reason why we focus on these particular eigenstates is that they are universal and the only ones that are relevant when taking large powers of the transfer matrices. As warm-up, we study the $k = 1$ and $k = 2$ cases, which have previously been worked out in the literature \cite{2020ScPP....8...67B}.

\subsection{k = 1}

For convenience, we reproduce the $k = 1$ (unreplicated) transfer matrix
\begin{equation}
 T^{(1)}_n = \frac{1}{q}
 \times 
 \scalebox{.7}{
 \begin{tikzpicture}[baseline=-3.5ex]
 \tblock{0}{0}{0};
 \draw[very thick] ({-1.03},{\rt}) -- (-1.03,-3) -- (15.03,-3) -- (15.03,{\rt}) ;
 \end{tikzpicture}}.
\end{equation}
Using unitarity \eqref{eq:unit}, we have the following relation
\begin{multline}
 T^{(1)}_n =
 \frac{1}{q}
 \times \scalebox{.7}{
 \begin{tikzpicture}[baseline=-3.5ex]
 \tblock{0}{0}{0};
 \draw[very thick] ({-1.03},{\rt}) -- (-1.03,-3) -- (15.03,-3) -- (15.03,{\rt}) ;
 \draw[very thick ] (6,-1.5)--(6,-2.25) -- (8,-2.25) -- (8,-1.5) ;
 \end{tikzpicture}}
 \\
 = \frac{1}{q}
 \times \scalebox{.7}{
 \begin{tikzpicture}[baseline=-3.5ex]
 \tblocknomid{0}{0}{0};
 \draw[very thick] ({-1.03},{\rt}) -- (-1.03,-3) -- (15.03,-3) -- (15.03,{\rt}) ;
 \draw[very thick ] (6,-1.5)--(6,-2.25) -- (8,-2.25) -- (8,-1.5) ;
 \end{tikzpicture}}.
\end{multline}
Contracting the two-site vector annihilates the neighboring unitaries.
Using dual-unitarity \eqref{eq:dualunit}, we may similarly annihilate neighboring unitaries in the other order
\begin{multline}
 T^{(1)}_n =
 \frac{1}{q}
 \times \scalebox{.7}{
 \begin{tikzpicture}[baseline=-3.5ex]
 \tblock{0}{0}{0};
 \draw[very thick] ({-1.03},{\rt}) -- (-1.03,-3) -- (15.03,-3) -- (15.03,{\rt}) ;
 \draw[very thick ] (0,-1.5)--(0,-2.25) -- (14,-2.25) -- (14,-1.5) ;
 \end{tikzpicture}}
 \\
 = \frac{1}{q}
 \times \scalebox{.7}{
 \begin{tikzpicture}[baseline=-3.5ex]
 \tblocknoout{0}{0}{0};
 \draw[very thick] ({-1.03},{\rt}) -- (-1.03,-3) -- (15.03,-3) -- (15.03,{\rt}) ;
 \draw[very thick ] (0,-1.5)--(0,-2.25) -- (14,-2.25) -- (14,-1.5) ;
 \end{tikzpicture}}.
\end{multline}
Therefore, in order to construct the eigenstates, we must contract neighboring red and blue operators until all operators are annihilated. At the point, the remaining loop contributes a factor of $q$, cancelling the $1/q$ in the transfer matrix.

It is useful to introduce a shorthand for enumerating such eigenstates where we remove the legs of the unitaries and replace the boxes by dots. Then, each eigenstate is given by connecting every red dot to a blue dot with a line, with none of the lines crossing each other. If the lines were to cross, this would correspond to annihilating unitaries that are not next to one another. For $T_n^{(1)}$ there is only a single way to contract all dots without crossings
\begin{equation}
 \begin{tikzpicture}
 \wick{2}{2.5}{1}
 \wick{1.5}{3}{2}
 \wick{.5}{4}{3}
 \wick{0}{4.5}{4}
 \dotchunk{0}
 \end{tikzpicture}
\end{equation}
Under the assumption that all eigenstates of eigenvalue one arise from these contractions, we conclude that
\begin{align}
 \lim_{m\rightarrow \infty} \left(T_n^{(1)}\right)^m = \begin{tikzpicture}[baseline=-7ex]
 \wick{2}{2.5}{1}
 \wick{1.5}{3}{2}
 \wick{.5}{4}{3}
 \wick{0}{4.5}{4}
 \wickflip{2}{2.5}{1}{0}{-2.5}
 \wickflip{1.5}{3}{2}{0}{-2.5}
 \wickflip{.5}{4}{3}{0}{-2.5}
 \wickflip{0}{4.5}{4}{0}{-2.5}
 \node[] at (1,0) {$\dots$};
 \node[] at (3.5,0) {$\dots$};
 \node[] at (1,-2.5) {$\dots$};
 \node[] at (3.5,-2.5) {$\dots$};
 \end{tikzpicture}.
\end{align}

\subsection{k = 2}
At $k = 2$, the transfer matrix is
\begin{equation}
 T^{(2)}_n =\frac{1}{q}
 \times 
 \scalebox{.4}{
 \begin{tikzpicture}[baseline=-3.5ex]
 \tblock{0}{0}{0};
 \tblock{0}{16}{0};
 \draw[very thick] ({-1.03},{\rt}) -- (-1.03,-3) -- (31.03,-3) -- (31.03,{\rt}) ;
 \end{tikzpicture}}.
\end{equation}
There are now several distinct eigenstates. To systematically analyze these, we start from the far left of the shorthand diagram. We observe that this far left dot is associated with the leftmost red dots and rightmost blue dots in each ``chunk''
\begin{equation}
 \begin{tikzpicture}
 \filldraw[orange] (0,0) circle (5pt);
 \filldraw[orange] (4.5,0) circle (5pt);
 \filldraw[orange] (5,0) circle (5pt);
 \filldraw[orange] (9.5,0) circle (5pt);
 \dotchunk{0}
 \dotchunk{5}
 \end{tikzpicture}.
\end{equation}
These dots are associated in the sense that they must be paired together because all other choices trap an unequal number of red and blue dots between them, which would force the lines to cross once all dots were contracted. 

There are two possibilities,
\begin{equation}
 \begin{tikzpicture}
 \wick{0}{4.5}{1}
 \wick{5}{9.5}{1}
 \dotchunk{0}
 \dotchunk{5}
 \end{tikzpicture},
\end{equation}
which we refer to as $\mathbbm{1}$ because it connects the first (second) red chunk with the first (second) blue chunk
and
\begin{equation}
 \begin{tikzpicture}
 \wick{0}{9.5}{2}
 \wick{4.5}{5}{1}
 \dotchunk{0}
 \dotchunk{5}
 \end{tikzpicture},
\end{equation}
which we refer to as $\mathbb{Z}_2$ because it connects swapped chunks.

We may now move inwards, where another set of four dots are associated
\begin{equation}
 \begin{tikzpicture}
 \filldraw[orange] (.5,0) circle (5pt);
 \filldraw[orange] (4,0) circle (5pt);
 \filldraw[orange] (5.5,0) circle (5pt);
 \filldraw[orange] (9,0) circle (5pt);
 \dotchunk{0}
 \dotchunk{5}
 \end{tikzpicture}.
\end{equation}
In the case where the first dot was chosen to be $\mathbbm{1}$, the second is also forced to be $\mathbbm{1}$ to avoid crossing
\begin{equation}
 \begin{tikzpicture}
 \wick{0}{4.5}{2}
 \wick{0.5}{4}{1}
 \wick{5}{9.5}{2}
 \wick{5.5}{9}{1}
 \dotchunk{0}
 \dotchunk{5}
 \end{tikzpicture}.
\end{equation}
If instead the first dot was chosen to be the $\mathbb{Z}_2$, then there remain the analogous two $\mathbbm{1}$ and $\mathbb{Z}_2$ choices,
\begin{equation}
 \begin{tikzpicture}
 \wick{0}{9.5}{2}
 \wick{.5}{4}{1}
 \wick{5.5}{9}{1}
 \wick{4.5}{5}{1}
 \dotchunk{0}
 \dotchunk{5}
 \end{tikzpicture}
\end{equation}
and
\begin{equation}
 \begin{tikzpicture}
 \wick{0}{9.5}{4}
 \wick{.5}{9}{3}
 \wick{4.5}{5}{1}
 \wick{4}{5.5}{2}
 \dotchunk{0}
 \dotchunk{5}
 \end{tikzpicture}.
\end{equation}
It should now be clear that this pattern continues. Once a set of four associated dots are contracted according to $\mathbbm{1}$, the remaining dots are forced to be contracted according to $\mathbbm{1}$ to avoid crossings. There are $n$ cases for this to occur and one case where all dots are contracted via $\mathbb{Z}_2$, leading to $n + 1$ eigenstates of unit eigenvalue.

This construction was somewhat straightforward, though will not immediately generalize to larger $k$. For this reason, we introduce what may initially appear to be a more complicated enumeration of the eigenstates, the advantage being that this enumeration will generalize. 
$\mathbb{Z}_2$ and $\mathbbm{1}$ comprise the group of permutations of two elements, $S_2$. 
Each eigenstate can then be labeled by a group element assigned to each set of associated dots
\begin{align}
 \ket{\tau_1, \tau_2, \dots \tau_n}, \quad \tau_i \in S_2.
\end{align}
It is important to note that not all of the above strings of permutations are legal eigenstates, namely $\tau_i$ cannot be $\mathbb{Z}_2$ if $\tau_j$ is $\mathbbm{1}$ for $i> j$. As permutations, this can be written as $\tau_i \leq \tau_j$ for $i > j$, where the ordering of the set is given by the order on the (very simple) lattice of permutations in $S_2$. We thus enumerate the states as
\begin{align}
 \ket{\tau_1 \geq \tau_2 \geq \dots \geq \tau_n}, \quad \tau_i \in S_2,
\end{align}
all of which are legal eigenstates.

In order to determine $\lim_{m\rightarrow \infty} (T_n^{(2)})^m$, we must orthonormalize the eigenstates. We can do this via the Gram-Schmidt procedure. We claim that the final result for the orthonormal vectors $\ket{i}$, $i\in \{0,1,\dots, $n$ \}$, is
\begin{equation}
 \begin{aligned}
& \ket{0} = \frac{1}{q^n}\, \ket{\mathbb{Z}_{2,1} \dots \mathbb{Z}_{2,n}}, \\
& \ket{i}=\frac{1}{q^n} \frac{1}{\sqrt{q^2-1}}\left(q \ket{\mathbb{Z}_{2,1} \dots \mathbb{Z}_{2,i}\mathbbm{1}_{i+1}\dots\mathbbm{1}_{n}}-\ket{\mathbb{Z}_{2,1} \dots \mathbb{Z}_{2,i-1}\mathbbm{1}_{i}\dots\mathbbm{1}_{n}}\right).
\end{aligned}
\end{equation} 
In order to check this result, we first note that
\begin{equation}
\bra{\mathbb{Z}_{2,1} \dots \mathbb{Z}_{2,l}\mathbbm{1}_{l+1}\dots\mathbbm{1}_{n}}{\mathbb{Z}_{2,1} \dots \mathbb{Z}_{2,m}\mathbbm{1}_{m+1}\dots\mathbbm{1}_{n}}\rangle = q^{2n-|l-m|}.
\end{equation}
The more general result is that
\begin{align}
\label{eq:overlapgen}
 \bra{\tilde{\tau}_1 \geq \dots \geq \tilde{\tau}_n}\tau_1 \geq \dots \geq \tau_n\rangle = q^{\sum_i C(\tilde{\tau}_i \tau_i^{-1})}.
\end{align}
where $C(\cdot)$ is the function that counts the number of cycles in the permutation, which is linearly related to the rank function. Using this overlap, we have for $i,j > 0$
\begin{align}
\begin{aligned}
 \bra{i}j\rangle = \frac{1}{q^{2n}} \frac{1}{{q^2-1}}\left(q^2q^{2n-|i-j|}-qq^{2n-|(i-1)-j|}-qq^{2n-|i-(j-1)|} +q^{2n-|(i-1)-(j-1)|}\right) = \delta_{ij}
 \\
 \bra{0}i\rangle = \frac{1}{q^{2n}} \frac{1}{\sqrt{q^2-1}}(qq^{n-i} - q^{n-(i-1)}) = 0, \quad \bra{0}0\rangle = \frac{1}{q^{2n}}q^{2n} = 1.
\end{aligned}
\end{align}
In the Graham-Schmidt process, it was necessary to first choose an order on the non-orthogonal basis vectors, which was naturally provided. We conclude that
\begin{align}
\label{eq:proj2}
\lim_{m\rightarrow \infty} (T_n^{(2)})^m = \sum_{i = 0}^n\ket{n}\bra{n}.
\end{align}

\subsection{General k}
We next recall the general replica transfer matrix
\begin{equation}
T^{(k)}_n =\frac{1}{q}
 \times 
\scalebox{0.2}{\begin{tikzpicture}[baseline=-6ex]
 \tblockshortflip{0}{0}{0};
 \tblockshortflip{0}{12}{0};
 \tblockshortflip{0}{24+18}{0};
 \tblockshortflip{0}{36+18}{0};
 \draw[very thick] (-1,{\rt}) -- (-1,{-.25*11-1.5}) -- (10+36+18+1, {-.25*11-1.5}) -- (10+36+18+1, {\rt});
 \node[scale = 3] at (32,-0.5) { $\textbf{\dots}$};
 \end{tikzpicture}}.
\end{equation}
In analogy with the $k = 2$ case of the prior subsection, out of the $2kn$ dots, there are $n$ sets of $2k$ dots associated with each other. These are the $j^{th}$ leftmost (rightmost) red (blue) dots in each chunk. Any other choice leads to a trapping of unequal numbers of red and blue dots. To be systematic, we start from the outside
\begin{equation}
\scalebox{.9}{
 \begin{tikzpicture}
 \filldraw[orange] (0,0) circle (5pt);
 \filldraw[orange] (4.5,0) circle (5pt);
 \filldraw[orange] (5,0) circle (5pt);
 \filldraw[orange] (9.5,0) circle (5pt);
 \filldraw[orange] (11.5,0) circle (5pt);
 \filldraw[orange] (16,0) circle (5pt);
 \dotchunk{0}
 \dotchunk{5}
 \dotchunk{11.5}
 \node[scale = 1.5] at (10.5,0) { $\textbf{\dots}$};
 \end{tikzpicture}}.
\end{equation}
Unlike the $k=2$ case, even for this first set, not all permutations in $S_k$ are allowed. For example, 
\begin{equation}
\scalebox{.9}{
 \begin{tikzpicture}
 \wick{0}{9.5}{2}
 \wick{4.5}{11.5}{3}
 \wick{5}{16}{1}
 \dotchunk{0}
 \dotchunk{5}
 \dotchunk{11.5}
 \node[scale = 1.5] at (10.5,0) { $\textbf{\dots}$};
 \end{tikzpicture}}
\end{equation}
has illegal crossings. It is thus clear that we must restrict to the noncrossing permutations, $NC_k$, such as
\begin{equation}
\scalebox{.9}{
 \begin{tikzpicture}
 \wick{0}{9.5}{2}
 \wick{4.5}{5}{1}
 \wick{11.5}{16}{1}
 \dotchunk{0}
 \dotchunk{5}
 \dotchunk{11.5}
 \node[scale = 1.5] at (10.5,0) { $\textbf{\dots}$};
 \end{tikzpicture}}.
\end{equation}
We may now move inwards
\begin{equation}
\scalebox{.9}{
 \begin{tikzpicture}
 \wick{0}{9.5}{2}
 \wick{4.5}{5}{1}
 \wick{11.5}{16}{1}
 \filldraw[orange] (0.5,0) circle (5pt);
 \filldraw[orange] (4,0) circle (5pt);
 \filldraw[orange] (5.5,0) circle (5pt);
 \filldraw[orange] (9,0) circle (5pt);
 \filldraw[orange] (12,0) circle (5pt);
 \filldraw[orange] (15.5,0) circle (5pt);
 \dotchunk{0}
 \dotchunk{5}
 \dotchunk{11.5}
 \node[scale = 1.5] at (10.5,0) { $\textbf{\dots}$};
 \end{tikzpicture}}
\end{equation}
We see that $\tau_1 \in NC_k$ places restrictions on the permutation $\tau_2$, the permutation connecting the orange dots above. $\tau_2$ must be a (noncrossing) permutation that is contained inside $\tau_1$ due to this constraint, i.e.~$\tau_2 \leq \tau_1$. Otherwise, the contractions from $\tau_2$ will cross the contractions of $\tau_1$. Applying this argument recursively, we determine that $\tau_i \leq \tau_{i-1}$ to avoid crossings and each $\tau_i$ must be noncrossing itself. This exhausts the list of eigenstates of unit eigenvalue for general $k$ and $n$ that can be determined only using unitary and dual unitarity. A specific model may have additional unit eigenvalue eigenstates but these require further structure on the dynamics. We thus enumerate the relevant eigenstates
by $n$-chains in the lattice of noncrossing partitions $NC_k$
\begin{align}
 \ket{\tau_1 \geq \dots \geq \tau_n}.
\end{align}
The total number of eigenstates is given by the Fuss-Catalan numbers \eqref{eq:FC}.
It is straightforward to check that, for $k = 1$ and $k=2$, the single and $n + 1$ eigenstates of the previous subsections are recovered.

In order to solve analytically solve for $\lim_{m\rightarrow \infty}\left(T_n^{(k)}\right)^m$, we need to orthonormalize the eigenvectors. This is challenging to achieve via a Graham-Schmidt procedure in generality, in part due to there not being a canonical ordering of $n$-chains, only a partial ordering. For any finite value of $n$ and $k$, an arbitrary ordering may be chosen and the Graham-Schmidt procedure may be carried out explicitly. For convenience, we now take the large-$q$ limit. 
It is clear from \eqref{eq:overlapgen} that in this limit, the eigenstates become asymptotically orthogonal. Normalizing this orthonormal basis gives
\begin{align}
 \frac{1}{q^{\frac{kn}{2}}} \ket{\tau_1 \geq \dots \geq \tau_n}, \quad \tau_i \in NC_k
\end{align}
and 
\begin{align}
\label{eq:projk}
\lim_{m\rightarrow \infty}\left(T_n^{(k)}\right)^m = \frac{1}{q^{{kn}}} \sum_{n-chains}\ket{\tau_1 \geq \dots \geq \tau_n}\bra{\tau_1 \geq \dots \geq \tau_n}.
\end{align}

\section{Out-of-time-ordered Correlators}

\label{sec:otoc}

\subsection{$k = 2$}

Due to the fact that each unitary acts on two sites, there is a parity effect in the OTOC. To retain generality, we place operators on adjacent sites. When specifying to a particular parity sector, we simply choose two of the operator insertions, e.g.~the orange and purple dots below, to be identity operators
\begin{align}
 {\scalebox{.3}{\begin{tikzpicture}
 \node[scale=3.3] at (-5,-5) {$C_{\alpha\gamma}^{(2)}(x,t) \propto $};
 \node[scale=3.3] at (14,-5) {$\propto$};
 \node[scale=3.3] at (24.5,-5) {$=$};
 \U{0}{0}{0} \U{0}{1.5}{1.5} 
 \U{0}{3}{0} \U{0}{4.5}{1.5} 
 \U{0}{6}{0} \U{0}{7.5}{1.5} 
 \U{0}{9}{0} \U{0}{10.5}{1.5} 
 \U{0}{0}{3} \U{0}{1.5}{4.5} 
 \U{0}{3}{3} \U{0}{4.5}{4.5} 
 \U{0}{6}{3} \U{0}{7.5}{4.5} 
 \U{0}{9}{3} \U{0}{10.5}{4.5} 
 \Ud{0}{1.5}{-6} \Ud{0}{3}{-4.5} 
 \Ud{0}{4.5}{-6} \Ud{0}{6}{-4.5} 
 \Ud{0}{7.5}{-6} \Ud{0}{9}{-4.5} 
 \Ud{0}{10.5}{-6} \Ud{0}{0}{-4.5} 
 \Ud{0}{1.5}{-3} \Ud{0}{3}{-1.5} 
 \Ud{0}{4.5}{-3} \Ud{0}{6}{-1.5} 
 \Ud{0}{7.5}{-3} \Ud{0}{9}{-1.5} 
 \Ud{0}{10.5}{-3} \Ud{0}{0}{-1.5} 
 \U{0}{0}{-12} \U{0}{1.5}{-10.5} 
 \U{0}{3}{-12} \U{0}{4.5}{-10.5} 
 \U{0}{6}{-12} \U{0}{7.5}{-10.5} 
 \U{0}{9}{-12} \U{0}{10.5}{-10.5} 
 \U{0}{0}{-9} \U{0}{1.5}{-7.5} 
 \U{0}{3}{-9} \U{0}{4.5}{-7.5} 
 \U{0}{6}{-9} \U{0}{7.5}{-7.5} 
 \U{0}{9}{-9} \U{0}{10.5}{-7.5} 
 \Ud{0}{1.5}{-18} \Ud{0}{3}{-16.5} 
 \Ud{0}{4.5}{-18} \Ud{0}{6}{-16.5} 
 \Ud{0}{7.5}{-18} \Ud{0}{9}{-16.5} 
 \Ud{0}{10.5}{-18} \Ud{0}{0}{-16.5} 
 \Ud{0}{1.5}{-15} \Ud{0}{3}{-13.5} 
 \Ud{0}{4.5}{-15} \Ud{0}{6}{-13.5} 
 \Ud{0}{7.5}{-15} \Ud{0}{9}{-13.5} 
 \Ud{0}{10.5}{-15} \Ud{0}{0}{-13.5} 
 \U{0}{16.5}{-10.5} 
 \U{0}{3+15}{-12} \U{0}{19.5}{-10.5} 
 \U{0}{3+15}{-9} \U{0}{19.5}{-7.5} 
 \U{0}{6+15}{-9} 
 \Ud{0}{18}{-16.5} 
 \Ud{0}{19.5}{-18} \Ud{0}{21}{-16.5} 
 \Ud{0}{16.5}{-15} \Ud{0}{18}{-13.5} 
 \Ud{0}{19.5}{-15} 
 \U{0}{16.5}{1.5} \U{0}{3+15}{0} \U{0}{19.5}{1.5} 
 \U{0}{3+15}{3} \U{0}{19.5}{4.5} \U{0}{6+15}{3} 
 \Ud{0}{18}{-4.5} \Ud{0}{19.5}{-6} \Ud{0}{21}{-4.5} 
 \Ud{0}{16.5}{-3} \Ud{0}{18}{-1.5} \Ud{0}{19.5}{-3} 
 \draw[very thick] (16.25,1.25)--(16.25,-1.75);
 \draw[very thick] (16.25,-9.25)--(16.25,1.5-4.75);
 \draw[very thick] (16.25,-18.25)--(16.25,1.5-16.75);
 \draw[very thick] (17.75,-18.25)--(17.75,-16.75);
 \draw[very thick] (17.75,1.25-9)--(17.75,-4.75);
 \draw[very thick] (22.25,-18.25)--(22.25,-16.75);
 \draw[very thick] (22.25,1.25-9)--(22.25,-4.75);
 \draw[very thick] (20.75,1.25)--(20.75,-1.75);
 \draw[very thick] (22.25,2.75)--(22.25,1.5-4.75);
 \draw[very thick] (22.25,-9.25)--(22.25,1.5-16.75);
 \draw[very thick] (16.25,-10.75)--(16.25,-13.75);
 \draw[very thick] (20.75,-10.75)--(20.75,-13.75);
 \draw[very thick] (22.25,4.25)--(22.25,5.75);
 \draw[very thick] (17.75,4.25)--(17.75,5.75);
 \draw[very thick] (16.25,2.75)--(16.25,5.75);
 \filldraw[orange] (17.75,-.25) circle (5pt);
 \draw[very thick] (17.75,-.25) circle (5pt);
 \filldraw[green] (20.75,-6.25) circle (5pt);
 \draw[very thick] (20.75,-6.25) circle (5pt);
 \filldraw[orange] (2.75,-.25) circle (5pt);
 \draw[very thick] (2.75,-.25) circle (5pt);
 \filldraw[green] (5.75,-6.25) circle (5pt);
 \draw[very thick] (5.75,-6.25) circle (5pt);
 \filldraw[orange] (17.75,-12.25) circle (5pt);
 \draw[very thick] (17.75,-12.25) circle (5pt);
 \filldraw[green] (20.75,-18.25) circle (5pt);
 \draw[very thick] (20.75,-18.25) circle (5pt);
 \filldraw[orange] (2.75,-12.25) circle (5pt);
 \draw[very thick] (2.75,-12.25) circle (5pt);
 \filldraw[green] (5.75,-18.25) circle (5pt);
 \draw[very thick] (5.75,-18.25) circle (5pt);
 \filldraw[black] (17.75+1.5,-.25) circle (5pt);
 \draw[very thick] (17.75+1.5,-.25) circle (5pt);
 \filldraw[purple] (20.75-1.5,-6.25) circle (5pt);
 \draw[very thick] (20.75-1.5,-6.25) circle (5pt);
 \filldraw[black] (2.75+1.5,-.25) circle (5pt);
 \draw[very thick] (2.75+1.5,-.25) circle (5pt);
 \filldraw[purple] (5.75-1.5,-6.25) circle (5pt);
 \draw[very thick] (5.75-1.5,-6.25) circle (5pt);
 \filldraw[black] (17.75+1.5,-12.25) circle (5pt);
 \draw[very thick] (17.75+1.5,-12.25) circle (5pt);
 \filldraw[purple] (20.75-1.5,-18.25) circle (5pt);
 \draw[very thick] (20.75-1.5,-18.25) circle (5pt);
 \filldraw[black] (2.75+1.5,-12.25) circle (5pt);
 \draw[very thick] (2.75+1.5,-12.25) circle (5pt);
 \filldraw[purple] (5.75-1.5,-18.25) circle (5pt);
 \draw[very thick] (5.75-1.5,-18.25) circle (5pt);
 \U{-45}{28}{2}
 \U{-45}{30}{2}
 \U{-45}{32}{2}
 \U{-45}{28}{0}
 \U{-45}{30}{0}
 \U{-45}{32}{0}
 \Ud{45}{{24-\rt+4}}{{-2+\rt}}
 \Ud{45}{{26-\rt+4}}{{-2+\rt}}
 \Ud{45}{{28-\rt+4}}{{-2+\rt}}
 \Ud{45}{{24-\rt+4}}{{-4+\rt}}
 \Ud{45}{{26-\rt+4}}{{-4+\rt}}
 \Ud{45}{{28-\rt+4}}{{-4+\rt}}
 \U{-45}{28}{2-8}
 \U{-45}{30}{2-8}
 \U{-45}{32}{2-8}
 \U{-45}{28}{0-8}
 \U{-45}{30}{0-8}
 \U{-45}{32}{0-8}
 \Ud{45}{{24-\rt+4}}{{-2+\rt-8}}
 \Ud{45}{{26-\rt+4}}{{-2+\rt-8}}
 \Ud{45}{{28-\rt+4}}{{-2+\rt-8}}
 \Ud{45}{{24-\rt+4}}{{-4+\rt-8}}
 \Ud{45}{{26-\rt+4}}{{-4+\rt-8}}
 \Ud{45}{{28-\rt+4}}{{-4+\rt-8}}
 \draw[very thick]({24+\rt/2+4},-2) -- ({24+\rt/2+4},0);
 \draw[very thick]({24+\rt/2+4},-10) -- ({24+\rt/2+4},0-8);
 \draw[very thick]({24+\rt/2+4},-4) --({24+\rt/2-1+4},-4) -- ({24+\rt/2-1+4},2)-- ({24+\rt/2+4},2);
 \draw[very thick]({24+\rt/2+4},-4-8) --({24+\rt/2-1+4},-4-8) -- ({24+\rt/2-1+4},2-8)-- ({24+\rt/2+4},2-8);
 \draw[very thick]({33.75},+2) -- ({33.75},3);
 \draw[very thick]({33.75},-6) -- ({33.75},0-4);
 \draw[very thick]({33.75},-6-7) -- ({33.75},0-4-8);
 \draw[very thick]({33.75},0) --({34.75},0) -- ({34.75},3);
 \draw[very thick]({33.75},-6-2) -- ({34.75},-6-2)-- ({34.75},0-2) -- ({33.75},0-2);
 \draw[very thick]({34.75},-13)-- ({34.75},0-10) -- ({33.75},0-10);
 \filldraw[green] ({33.75},-6+1) circle (5pt);
 \draw[very thick] ({33.75},-6+1) circle (5pt);
 \filldraw[green] ({33.75},-6-8+1) circle (5pt);
 \draw[very thick] ({33.75},-6-8+1) circle (5pt);
 \filldraw[orange] ({24+\rt/2+4},-9) circle (5pt);
 \draw[very thick] ({24+\rt/2+4},-9) circle (5pt);
 \filldraw[orange] ({24+\rt/2+4},-1) circle (5pt);
 \draw[very thick] ({24+\rt/2+4},-1) circle (5pt);
 \filldraw[purple] ({33.75-1.05},-6+1) circle (5pt);
 \draw[very thick] ({33.75-1.05},-6+1) circle (5pt);
 \filldraw[purple] ({33.75-1.05},-6-8+1) circle (5pt);
 \draw[very thick] ({33.75-1.05},-6-8+1) circle (5pt);
 \filldraw[black] ({25.05+\rt/2+4},-9) circle (5pt);
 \draw[very thick] ({25.05+\rt/2+4},-9) circle (5pt);
 \filldraw[black] ({25.05+\rt/2+4},-1) circle (5pt);
 \draw[very thick] ({25.05+\rt/2+4},-1) circle (5pt);
 \end{tikzpicture}}}.
\end{align}
In going from the first diagram to the second, we have used the unitarity of the gates. In going from the second to the third, we have simply compressed the diagram. For simplicity, we took $t = 4$ and $x \in \{0,1,2\}$, depending on the parity chosen, though the structure clearly generalizes. The trace, connecting the top and bottom edges, is implicit. In the final diagram, the columns are seen to be the $k = 2$ and $n = (t-x + c_n)/2$ transfer matrix with $c_n \in \{ 0,1,2\}$ depending on the parity. We thus find that the four-point OTOC is given by
\begin{align}
 \langle\sigma_{\alpha}(0,0) \sigma_{\gamma}(x,t) \sigma_{\alpha}(0,0) \sigma_{\gamma}(x,t)\rangle = \bra{L_{i}^{(2)}(\sigma_{\alpha})} \left(T_n^{(2)}\right)^m \ket{R_{i}^{(2)}(\sigma_{\gamma})}
\end{align}
where $m = (x + t + c_m)/2$ with $c_m \in \{0, 1, 2\}$ depending on parity. The boundary conditions also depend on parity. When the orange (green) dot, denoting $\sigma_\alpha$ ($\sigma_\gamma$) on the left (right), is nontrivial but the black (purple) dot is trivial, we write a subscript $1$ for the boundary condition. The opposite configuration is given by the subscript $2$
\begin{align}
 \bra{L_1^{(2)}(\sigma_\alpha)} \;\;&\equiv \frac{1}{q^{n/2}}\quad\scalebox{1}{\begin{tikzpicture}[baseline=.5ex]
 \wickflip{1}{2}{1}{0}{0}
 \wickflip{0}{3}{2}{0}{0}
 \wickflip{1+4}{2+4}{1}{0}{0}
 \wickflip{0+4}{3+4}{2}{0}{0}
 \filldraw[orange] (1.5,.25) circle (2pt);
 \draw[ thick] (1.5,.25) circle (2pt);
 \filldraw[orange] (1.5+4,.25) circle (2pt);
 \draw[ thick] (1.5+4,.25) circle (2pt);
 \node[scale = .5] at (.5,0) { $\textbf{\dots}$};
 \node[scale = .5] at (2.5,0) { $\textbf{\dots}$};
 \node[scale = .5] at (.5+4,0) { $\textbf{\dots}$};
 \node[scale = .5] at (2.5+4,0) { $\textbf{\dots}$};
 \draw [decorate,decoration={brace,amplitude=2pt,mirror,raise=2pt}]
 (0,-.1) -- (1,-.1) node[midway,yshift=-9pt,scale=.5]{n};
 \end{tikzpicture}},
 \nonumber\\
 \bra{L^{(2)}_2(\sigma_\alpha)} \;\;&\equiv \frac{1}{q^{n/2+1}} \;\;
 \scalebox{0.4}{\begin{tikzpicture}[baseline=-6ex]
 \tblockshortflip{0}{0}{0};
 \tblockshortflip{0}{12}{0};
 \wickflipU{4}{6}{3};
 \wickflipU{0}{10}{5};
 \wickflipU{4+12}{6+12}{3};
 \wickflipU{12}{22}{5};
 \wickflipU{-1}{23}{7};
 \draw[very thick] (-1,0) -- (-1,-.7);
 \draw[very thick] (23,0) -- (23,-.7);
 \draw[thick] (5,{\rt}) circle (5pt);
 \filldraw[black] (5,{\rt}) circle (5pt);
 \draw[thick] (5+12,{\rt}) circle (5pt);
 \filldraw[black] (5+12,{\rt}) circle (5pt);
 \end{tikzpicture}},
 \nonumber\\
 \ket{R_1^{(2)}(\sigma_\gamma)} \;\;&\equiv \frac{1}{q^{n/2}}\quad\scalebox{1}{\begin{tikzpicture}[baseline=-1.5ex]
 \wick{-1}{4}{3}
 \wick{-2}{5}{4}
 \wick{1}{2}{1}
 \wick{0}{3}{2}
 \filldraw[green] (1.5,-.25) circle (2pt);
 \draw[ thick] (1.5,-.25) circle (2pt);
 \filldraw[green] (1.5,-1) circle (2pt);
 \draw[thick] (1.5,-1) circle (2pt);
 \node[scale = .5] at (.5,0) { $\textbf{\dots}$};
 \node[scale = .5] at (2.5,0) { $\textbf{\dots}$};
 \node[scale = .5] at (4.5,0) { $\textbf{\dots}$};
 \node[scale = .5] at (.5-2,0) { $\textbf{\dots}$};
 \node[scale = .5] at (2.5-2,0) { $\textbf{\dots}$};
 \end{tikzpicture}},\nonumber\\
 \ket{R_2^{(2)}(\sigma_\gamma)} \;\;&\equiv\frac{1}{q^{n/2+1}} \;\; \scalebox{0.4}{\begin{tikzpicture}[baseline=-6ex]
 \tblockshortflip{0}{0}{0};
 \tblockshortflip{0}{12}{0};
 \wickU{4}{18}{7};
 \wickU{0}{22}{9};
 \wickU{-1}{23}{11};
 \wickU{6}{16}{5};
 \wickU{10}{12}{3}; 
 \draw[very thick] (-1,-1.7) -- (-1,-.7);
 \draw[very thick] (23,-1.7) -- (23,-.7);
 \filldraw[purple] (11,{\rt}) circle (5pt);
 \draw[thick] (11,{\rt}) circle (5pt);
 \filldraw[purple] (23,{\rt}) circle (5pt);
 \draw[thick] (23,{\rt}) circle (5pt);
 \end{tikzpicture}}.
\end{align}


The cases with left boundary conditions $\bra{L_1^{(2)}(\sigma_{\alpha})}$ were studied in \cite{2020PhRvR...2c3032C} at finite $q$. Specifying to far along the light cone but a finite distance from the light cone (large $m$, $n$ finite), the dominant contribution to the OTOC comes from the projector \eqref{eq:proj2}, which at large $q$ becomes
\begin{align}
\lim_{m\rightarrow \infty} (T_n^{(2)})^m = \frac{1}{q^{2n}}\sum_{i = 0}^n\ket{\mathbb{Z}_{2,1} \dots \mathbb{Z}_{2,i}\mathbbm{1}_{i+1}\dots\mathbbm{1}_{n}}\bra{\mathbb{Z}_{2,1} \dots \mathbb{Z}_{2,i}\mathbbm{1}_{i+1}\dots\mathbbm{1}_{n}}.
\end{align}
The inner products of the left boundary condition and these basis states are
\begin{align}
\bra{L_1^{(2)}(\sigma_{\alpha})}\mathbb{Z}_{2,1} \dots \mathbb{Z}_{2,i}\mathbbm{1}_{i+1}\dots\mathbbm{1}_{n}\rangle =q^{n/2} \delta_{in} .
\end{align}
due to the tracelessness of the Pauli operators. For $\ket{R_1^{(2)}(\sigma_{\gamma})}$, this implies a trivial OTOC because
\begin{align}
 \bra{\mathbb{Z}_{2,1} \dots \mathbb{Z}_{2,n}}R_1^{(2)}(\sigma_{\gamma})\rangle = 0.
\end{align}
However, the other right boundary condition gives a nontrivial OTOC.
Using the channels defined previously \eqref{eq:channel1}, we have 
\begin{align}
\begin{aligned}
 C_{\alpha\gamma}^{(2)}(x,t)&=\frac{1}{q^{3n/2}}\bra{\mathbb{Z}_{2,1} \dots \mathbb{Z}_{2,n}}R_2^{(2)}(\sigma_{\gamma})\rangle
 \\
 &=\frac{1}{q^{2n+1}}\quad\scalebox{0.3}{\begin{tikzpicture}[baseline=-6ex]
 \tblockshortflip{0}{0}{0};
 \tblockshortflip{0}{12}{0};
 \wickU{0}{22}{9};
 \wickU{4}{18}{7};
 \wickU{6}{16}{5};
 \wickU{10}{12}{3};
 \wickflipU{0}{22}{9};
 \wickflipU{4}{18}{7};
 \wickflipU{6}{16}{5};
 \wickflipU{10}{12}{3};
 \draw[thick] (11,{\rt}) circle (3pt);
 \filldraw[purple] (11,{\rt}) circle (3pt);
 \draw[thick] (23,{\rt}) circle (3pt);
 \filldraw[purple] (23,{\rt}) circle (3pt);
 \node[scale = 1.7] at (11,-0.1) { $\sigma_\gamma$};
 \node[scale = 1.7] at (23,-0.1) { $\sigma_\gamma$};
 \draw[very thick] (-1,{\rt}) -- (-1,{-.25*11-1.5}) -- (23, {-.25*11-1.5}) -- (23, {\rt});
 \end{tikzpicture}}\\
 &= \frac{1}{q}\Tr((\mathcal{M}^n_{+}(\sigma_\gamma))^2)
 \end{aligned}.
\end{align} 
Due to the unistochasticity of $\mathcal{M}_+$, we find that the OTOC decays exponentially as $\sim \lambda_{max}^{-2n}$, where $\lambda_{max}$ is the largest nontrivial eigenvalue of the channel. By symmetry, the OTOC is the same for boundary conditions $\bra{L_2^{(2)}(\sigma_{\alpha})}$ and $\ket{R_1^{(2)}(\sigma_{\gamma})}$. The most complicated case, with boundary conditions $\bra{L_2^{(2)}(\sigma_{\alpha})}$ and $\ket{R_2^{(2)}(\sigma_{\gamma})}$, is treated in the following subsection and it turns out not to have a simple solution.



\subsection{General $k$}

The general $2k$-point OTOC is defined as $$C_{\alpha\gamma}^{(k)}(x,t)=\langle (\sigma_\alpha(0,0)\sigma_\gamma(x,t))^k\rangle$$ 
As an illustration, we use the same simple configuration of operators as in the previous subsection
\begin{align}
 C_{\alpha\gamma}^{(k)}(x,t) \propto \quad \raisebox{-150pt} {\scalebox{.3}{\begin{tikzpicture}
 \node[scale=3.3] at (14,-13) {$\propto$};
 \node[scale=3.3] at (24.5,-13) {$=$};
 \node[scale=4.5] at (6,-32) {$\vdots$};
 \node[scale=4.5] at (19,-32) {$\vdots$};
 \node[scale=4.5] at (30,-22) {$\vdots$};
 \draw[thick ,decoration={
 brace
 },decorate](-1,-6) -- (-1,5.8) 
node [pos=0.5,xshift=-3.3cm, yshift=0cm, scale=3] {$k$ of these}; 
 \U{0}{0}{0} \U{0}{1.5}{1.5} 
 \U{0}{3}{0} \U{0}{4.5}{1.5} 
 \U{0}{6}{0} \U{0}{7.5}{1.5} 
 \U{0}{9}{0} \U{0}{10.5}{1.5} 
 \U{0}{0}{3} \U{0}{1.5}{4.5} 
 \U{0}{3}{3} \U{0}{4.5}{4.5} 
 \U{0}{6}{3} \U{0}{7.5}{4.5} 
 \U{0}{9}{3} \U{0}{10.5}{4.5} 
 \Ud{0}{1.5}{-6} \Ud{0}{3}{-4.5} 
 \Ud{0}{4.5}{-6} \Ud{0}{6}{-4.5} 
 \Ud{0}{7.5}{-6} \Ud{0}{9}{-4.5} 
 \Ud{0}{10.5}{-6} \Ud{0}{0}{-4.5} 
 \Ud{0}{1.5}{-3} \Ud{0}{3}{-1.5} 
 \Ud{0}{4.5}{-3} \Ud{0}{6}{-1.5} 
 \Ud{0}{7.5}{-3} \Ud{0}{9}{-1.5} 
 \Ud{0}{10.5}{-3} \Ud{0}{0}{-1.5} 
 \U{0}{0}{-12} \U{0}{1.5}{-10.5} 
 \U{0}{3}{-12} \U{0}{4.5}{-10.5} 
 \U{0}{6}{-12} \U{0}{7.5}{-10.5} 
 \U{0}{9}{-12} \U{0}{10.5}{-10.5} 
 \U{0}{0}{-9} \U{0}{1.5}{-7.5} 
 \U{0}{3}{-9} \U{0}{4.5}{-7.5} 
 \U{0}{6}{-9} \U{0}{7.5}{-7.5} 
 \U{0}{9}{-9} \U{0}{10.5}{-7.5} 
 \Ud{0}{1.5}{-18} \Ud{0}{3}{-16.5} 
 \Ud{0}{4.5}{-18} \Ud{0}{6}{-16.5} 
 \Ud{0}{7.5}{-18} \Ud{0}{9}{-16.5} 
 \Ud{0}{10.5}{-18} \Ud{0}{0}{-16.5} 
 \Ud{0}{1.5}{-15} \Ud{0}{3}{-13.5} 
 \Ud{0}{4.5}{-15} \Ud{0}{6}{-13.5} 
 \Ud{0}{7.5}{-15} \Ud{0}{9}{-13.5} 
 \Ud{0}{10.5}{-15} \Ud{0}{0}{-13.5} 
 \U{0}{0}{-24} \U{0}{1.5}{-22.5} 
 \U{0}{3}{0-24} \U{0}{4.5}{-22.5} 
 \U{0}{6}{0-24} \U{0}{7.5}{-22.5} 
 \U{0}{9}{0-24} \U{0}{10.5}{-22.5} 
 \U{0}{0}{3-24} \U{0}{1.5}{4.5-24} 
 \U{0}{3}{3-24} \U{0}{4.5}{4.5-24} 
 \U{0}{6}{3-24} \U{0}{7.5}{4.5-24} 
 \U{0}{9}{3-24} \U{0}{10.5}{4.5-24} 
 \Ud{0}{1.5}{-6-24} \Ud{0}{3}{-28.5} 
 \Ud{0}{4.5}{-6-24} \Ud{0}{6}{-28.5} 
 \Ud{0}{7.5}{-6-24} \Ud{0}{9}{-28.5} 
 \Ud{0}{10.5}{-6-24} \Ud{0}{0}{-28.5} 
 \Ud{0}{1.5}{-27} \Ud{0}{3}{-25.5} 
 \Ud{0}{4.5}{-27} \Ud{0}{6}{-25.5} 
 \Ud{0}{7.5}{-27} \Ud{0}{9}{-25.5} 
 \Ud{0}{10.5}{-27} \Ud{0}{0}{-25.5} 
 \U{0}{16.5}{1.5} 
 \U{0}{3+15}{0} \U{0}{19.5}{1.5} 
 \U{0}{3+15}{3} \U{0}{19.5}{4.5} 
 \U{0}{6+15}{3} 
 \Ud{0}{18}{-4.5} 
 \Ud{0}{19.5}{-6} \Ud{0}{21}{-4.5} 
 \Ud{0}{16.5}{-3} \Ud{0}{18}{-1.5} 
 \Ud{0}{19.5}{-3} 
 \draw[very thick] (16.25,1.25)--(16.25,-1.75);
 \draw[very thick] (16.25,-9.25)--(16.25,1.5-4.75);
 \draw[very thick] (16.25,1.25-24+1.5)--(16.25,1.5-16.75);
 \draw[very thick] (16.25,1.25-31.5)--(16.25,-27.25);
 \draw[very thick] (17.75,1.25-30-1.5)--(17.75,-28.75);
 \draw[very thick] (17.75,1.25-21)--(17.75,-16.75);
 \draw[very thick] (17.75,1.25-9)--(17.75,-4.75);
 \draw[very thick] (22.25,1.25-21)--(22.25,-16.75);
 \draw[very thick] (22.25,1.25-9)--(22.25,-4.75);
 \draw[very thick] (20.75,1.25)--(20.75,-1.75);
 \draw[very thick] (22.25,2.75)--(22.25,1.5-4.75);
 \draw[very thick] (22.25,-9.25)--(22.25,1.5-16.75);
 \draw[very thick] (22.25,1.25+-22.5)--(22.25,-27.25);
 \draw[very thick] (16.25,-10.75)--(16.25,-13.75);
 \draw[very thick] (20.75,-10.75)--(20.75,-13.75);
 \draw[very thick] (16.25,1.25-24)--(16.25,-1.75-24);
 \draw[very thick] (20.75,1.25-24)--(20.75,-1.75-24);
 \draw[very thick] (22.25,4.25)--(22.25,5.75);
 \draw[very thick] (22.25,1.25-31.5)--(22.25,-28.75);
 \draw[very thick] (17.75,4.25)--(17.75,5.75);
 \draw[very thick] (16.25,2.75)--(16.25,5.75);
 \U{0}{16.5}{-10.5} 
 \U{0}{3+15}{-12} \U{0}{19.5}{-10.5} 
 \U{0}{3+15}{-9} \U{0}{19.5}{-7.5} 
 \U{0}{6+15}{-9} 
 \Ud{0}{18}{-16.5} 
 \Ud{0}{19.5}{-18} \Ud{0}{21}{-16.5} 
 \Ud{0}{16.5}{-15} \Ud{0}{18}{-13.5} 
 \Ud{0}{19.5}{-15} 
 \U{0}{16.5}{-22.5} 
 \U{0}{3+15}{0-24} \U{0}{19.5}{-22.5} 
 \U{0}{3+15}{3-24} \U{0}{19.5}{4.5-24} 
 \U{0}{6+15}{3-24} 
 \Ud{0}{18}{-28.5} 
 \Ud{0}{19.5}{-6-24} \Ud{0}{21}{-28.5} 
 \Ud{0}{16.5}{-27} \Ud{0}{18}{-25.5} 
 \Ud{0}{19.5}{-27} 
 \filldraw[orange] (17.75,-.25) circle (5pt);
 \draw[very thick] (17.75,-.25) circle (5pt);
 \filldraw[green] (20.75,-6.25) circle (5pt);
 \draw[very thick] (20.75,-6.25) circle (5pt);
 \filldraw[orange] (2.75,-.25) circle (5pt);
 \draw[very thick] (2.75,-.25) circle (5pt);
 \filldraw[green] (5.75,-6.25) circle (5pt);
 \draw[very thick] (5.75,-6.25) circle (5pt);
 \filldraw[orange] (17.75,-12.25) circle (5pt);
 \draw[very thick] (17.75,-12.25) circle (5pt);
 \filldraw[orange] (17.75,-24.25) circle (5pt);
 \draw[very thick] (17.75,-24.25) circle (5pt);
 \filldraw[green] (20.75,-18.25) circle (5pt);
 \draw[very thick] (20.75,-18.25) circle (5pt);
 \filldraw[green] (20.75,-30.25) circle (5pt);
 \draw[very thick] (20.75,-30.25) circle (5pt);
 \filldraw[orange] (2.75,-12.25) circle (5pt);
 \draw[very thick] (2.75,-12.25) circle (5pt);
 \filldraw[orange] (2.75,-24.25) circle (5pt);
 \draw[very thick] (2.75,-24.25) circle (5pt);
 \filldraw[green] (5.75,-18.25) circle (5pt);
 \draw[very thick] (5.75,-18.25) circle (5pt);
 \filldraw[green] (5.75,-30.25) circle (5pt);
 \draw[very thick] (5.75,-30.25) circle (5pt);
 \filldraw[black] (17.75+1.5,-.25) circle (5pt);
 \draw[very thick] (17.75+1.5,-.25) circle (5pt);
 \filldraw[purple] (20.75-1.5,-6.25) circle (5pt);
 \draw[very thick] (20.75-1.5,-6.25) circle (5pt);
 \filldraw[black] (2.75+1.5,-.25) circle (5pt);
 \draw[very thick] (2.75+1.5,-.25) circle (5pt);
 \filldraw[purple] (5.75-1.5,-6.25) circle (5pt);
 \draw[very thick] (5.75-1.5,-6.25) circle (5pt);
 \filldraw[black] (17.75+1.5,-12.25) circle (5pt);
 \draw[very thick] (17.75+1.5,-12.25) circle (5pt);
 \filldraw[black] (17.75+1.5,-24.25) circle (5pt);
 \draw[very thick] (17.75+1.5,-24.25) circle (5pt);
 \filldraw[purple] (20.75-1.5,-18.25) circle (5pt);
 \draw[very thick] (20.75-1.5,-18.25) circle (5pt);
 \filldraw[purple] (20.75-1.5,-30.25) circle (5pt);
 \draw[very thick] (20.75-1.5,-30.25) circle (5pt);
 \filldraw[black] (2.75+1.5,-12.25) circle (5pt);
 \draw[very thick] (2.75+1.5,-12.25) circle (5pt);
 \filldraw[black] (2.75+1.5,-24.25) circle (5pt);
 \draw[very thick] (2.75+1.5,-24.25) circle (5pt);
 \filldraw[purple] (5.75-1.5,-18.25) circle (5pt);
 \draw[very thick] (5.75-1.5,-18.25) circle (5pt);
 \filldraw[purple] (5.75-1.5,-30.25) circle (5pt);
 \draw[very thick] (5.75-1.5,-30.25) circle (5pt);
 \U{-45}{28}{2}
 \U{-45}{30}{2}
 \U{-45}{32}{2}
 \U{-45}{28}{0}
 \U{-45}{30}{0}
 \U{-45}{32}{0}
 \Ud{45}{{24-\rt+4}}{{-2+\rt}}
 \Ud{45}{{26-\rt+4}}{{-2+\rt}}
 \Ud{45}{{28-\rt+4}}{{-2+\rt}}
 \Ud{45}{{24-\rt+4}}{{-4+\rt}}
 \Ud{45}{{26-\rt+4}}{{-4+\rt}}
 \Ud{45}{{28-\rt+4}}{{-4+\rt}}
 \U{-45}{28}{2-8}
 \U{-45}{30}{2-8}
 \U{-45}{32}{2-8}
 \U{-45}{28}{0-8}
 \U{-45}{30}{0-8}
 \U{-45}{32}{0-8}
 \Ud{45}{{24-\rt+4}}{{-2+\rt-8}}
 \Ud{45}{{26-\rt+4}}{{-2+\rt-8}}
 \Ud{45}{{28-\rt+4}}{{-2+\rt-8}}
 \Ud{45}{{24-\rt+4}}{{-4+\rt-8}}
 \Ud{45}{{26-\rt+4}}{{-4+\rt-8}}
 \Ud{45}{{28-\rt+4}}{{-4+\rt-8}}
 \U{-45}{28}{2-16}
 \U{-45}{30}{2-16}
 \U{-45}{32}{2-16}
 \U{-45}{28}{0-16}
 \U{-45}{30}{0-16}
 \U{-45}{32}{0-16}
 \Ud{45}{{24-\rt+4}}{{-2+\rt-16}}
 \Ud{45}{{26-\rt+4}}{{-2+\rt-16}}
 \Ud{45}{{28-\rt+4}}{{-2+\rt-16}}
 \Ud{45}{{24-\rt+4}}{{-4+\rt-16}}
 \Ud{45}{{26-\rt+4}}{{-4+\rt-16}}
 \Ud{45}{{28-\rt+4}}{{-4+\rt-16}}
 \draw[very thick]({24+\rt/2+4},-2) -- ({24+\rt/2+4},0);
 \draw[very thick]({24+\rt/2+4},-10) -- ({24+\rt/2+4},0-8);
 \draw[very thick]({24+\rt/2+4},-2-16) -- ({24+\rt/2+4},0-16);
 \draw[very thick]({24+\rt/2+4},-4) --({24+\rt/2-1+4},-4) -- ({24+\rt/2-1+4},2)-- ({24+\rt/2+4},2);
 \draw[very thick]({24+\rt/2+4},-4-8) --({24+\rt/2-1+4},-4-8) -- ({24+\rt/2-1+4},2-8)-- ({24+\rt/2+4},2-8);
 \draw[very thick]({24+\rt/2+4},-4-16) --({24+\rt/2-1+4},-4-16) -- ({24+\rt/2-1+4},2-16)-- ({24+\rt/2+4},2-16);
 \draw[very thick]({33.75},+2) -- ({33.75},3);
 \draw[very thick]({33.75},-6) -- ({33.75},0-4);
 \draw[very thick]({33.75},-6-8) -- ({33.75},0-4-8);
 \draw[very thick]({33.75},-1-4-16) -- ({33.75},0-4-16);
 \draw[very thick]({33.75},0) --({34.75},0) -- ({34.75},3);
 \draw[very thick]({33.75},-6-2) -- ({34.75},-6-2)-- ({34.75},0-2) -- ({33.75},0-2);
 \draw[very thick]({33.75},-16) -- ({34.75},-16)-- ({34.75},0-10) -- ({33.75},0-10);
 \draw[very thick]({33.75},-18) -- ({34.75},-18) -- ({34.75},-21); 
 \filldraw[green] ({33.75},-6+1) circle (5pt);
 \draw[very thick] ({33.75},-6+1) circle (5pt);
 \filldraw[green] ({33.75},-6-8+1) circle (5pt);
 \draw[very thick] ({33.75},-6-8+1) circle (5pt);
 \filldraw[green] ({33.75},-6-16+1) circle (5pt);
 \draw[very thick] ({33.75},-6-16+1) circle (5pt);
 \filldraw[orange] ({24+\rt/2+4},-9) circle (5pt);
 \draw[very thick] ({24+\rt/2+4},-9) circle (5pt);
 \filldraw[orange] ({24+\rt/2+4},-17) circle (5pt);
 \draw[very thick] ({24+\rt/2+4},-17) circle (5pt);
 \filldraw[orange] ({24+\rt/2+4},-1) circle (5pt);
 \draw[very thick] ({24+\rt/2+4},-1) circle (5pt);
 \filldraw[purple] ({33.75-1.05},-6+1) circle (5pt);
 \draw[very thick] ({33.75-1.05},-6+1) circle (5pt);
 \filldraw[purple] ({33.75-1.05},-6-8+1) circle (5pt);
 \draw[very thick] ({33.75-1.05},-6-8+1) circle (5pt);
 \filldraw[purple] ({33.75-1.05},-6-16+1) circle (5pt);
 \draw[very thick] ({33.75-1.05},-6-16+1) circle (5pt);
 \filldraw[black] ({25.05+\rt/2+4},-9) circle (5pt);
 \draw[very thick] ({25.05+\rt/2+4},-9) circle (5pt);
 \filldraw[black] ({25.05+\rt/2+4},-1) circle (5pt);
 \draw[very thick] ({25.05+\rt/2+4},-1) circle (5pt);
 \filldraw[black] ({25.05+\rt/2+4},-17) circle (5pt);
 \draw[very thick] ({25.05+\rt/2+4},-17) circle (5pt);
 \end{tikzpicture}}}
\end{align}
The proportionality factor on the last diagram is again 
most easily determined by demanding that $C_{00}^{(k)}(x,t)=\langle \mathbbm{1}\rangle = 1$.
Each column in the final diagram is seen to be a transfer matrix $T^{(k)}_n $ \eqref{eq:tkn}. These transfer matrices are sandwiched between two boundary conditions states. 


The possible boundary conditions can be read off from the diagrams
\begin{align}
\begin{aligned}
 \bra{L_1^{(k)}(\sigma_\alpha)} \;\;&\equiv \frac{1}{q^{n/2}}\quad\scalebox{.55}{\begin{tikzpicture}[baseline=.5ex]
 \wickflip{1}{2}{1}{0}{0}
 \wickflip{0}{3}{2}{0}{0}
 \wickflip{1+4}{2+4}{1}{0}{0}
 \wickflip{0+4}{3+4}{2}{0}{0}
 \wickflip{1+18}{2+18}{1}{0}{0}
 \wickflip{0+18}{3+18}{2}{0}{0}
 \wickflip{1+14}{2+14}{1}{0}{0}
 \wickflip{0+14}{3+14}{2}{0}{0}
 \filldraw[orange] (1.5,.25) circle (3pt);
 \draw[ thick] (1.5,.25) circle (3pt);
 \filldraw[orange] (1.5+4,.25) circle (3pt);
 \draw[ thick] (1.5+4,.25) circle (3pt);
 \filldraw[orange] (1.5+18,.25) circle (3pt);
 \draw[thick] (1.5+18,.25) circle (3pt);
 \filldraw[orange] (1.5+14,.25) circle (3pt);
 \draw[thick] (1.5+14,.25) circle (3pt);
 \node[scale = 1.5] at (8.5+2,0) { $\textbf{\dots}$};
 \node[scale = .5] at (.5,0) { $\textbf{\dots}$};
 \node[scale = .5] at (2.5,0) { $\textbf{\dots}$};
 \node[scale = .5] at (.5+4,0) { $\textbf{\dots}$};
 \node[scale = .5] at (2.5+4,0) { $\textbf{\dots}$};
 \node[scale = .5] at (.5+14,0) { $\textbf{\dots}$};
 \node[scale = .5] at (2.5+14,0) { $\textbf{\dots}$};
 \node[scale = .5] at (.5+18,0) { $\textbf{\dots}$};
 \node[scale = .5] at (2.5+18,0) { $\textbf{\dots}$};
 \draw [decorate,decoration={brace,amplitude=2pt,mirror,raise=2pt}]
 (0,-.1) -- (1,-.1) node[midway,yshift=-9pt,scale=1]{$n$};
 \end{tikzpicture}}
 \\
 \bra{L_{2}^{(k)}(\sigma_\alpha)} \;\;&\equiv \frac{1}{q^{n/2+1}} \;\;
 \scalebox{0.18}{\begin{tikzpicture}[baseline=-6ex]
 \tblockshortflip{0}{0}{0};
 \tblockshortflip{0}{12}{0};
 \tblockshortflip{0}{24+18}{0};
 \tblockshortflip{0}{36+18}{0};
 \wickflipU{4}{6}{3};
 \wickflipU{0}{10}{5};
 \wickflipU{4+12}{6+12}{3};
 \wickflipU{0+12}{10+12}{5};
 \wickflipU{4+24+18}{6+24+18}{3};
 \wickflipU{0+24+18}{10+24+18}{5};
 \wickflipU{4+36+18}{6+36+18}{3};
 \wickflipU{0+36+18}{10+36+18}{5};
 \draw[very thick] (-1,{\rt}) -- (-1,{.25*7}) -- (10+36+18+1, {.25*7}) -- (10+36+18+1, {\rt});
 \node[scale = 3] at (32,-0.5) { $\textbf{\dots}$};
 \draw[thick] (5,{\rt}) circle (6pt);
 \filldraw[black] (5,{\rt}) circle (6pt);
 \draw[thick] (5+12,{\rt}) circle (6pt);
 \filldraw[black] (5+12,{\rt}) circle (6pt);
 \draw[thick] (5+24+18,{\rt}) circle (6pt);
 \filldraw[black] (5+24+18,{\rt}) circle (6pt);
 \draw[thick] (5+36+18,{\rt}) circle (6pt);
 \filldraw[black] (5+36+18,{\rt}) circle (6pt);
 \end{tikzpicture}}
 \\
 \ket{R^{(k)}_1(\sigma_\gamma)} \;\;&\equiv \frac{1}{q^{n/2}}\quad\scalebox{.55}{\begin{tikzpicture}[baseline=-1.5ex]
 \wick{-1}{18}{3}
 \wick{-2}{19}{4}
 \wick{1}{2}{1}
 \wick{0}{3}{2}
 \wick{1+4}{2+4}{1}
 \wick{0+4}{3+4}{2}
 \wick{1+10}{2+10}{1}
 \wick{0+10}{3+10}{2}
 \wick{1+14}{2+14}{1}
 \wick{0+14}{3+14}{2}
 \filldraw[green] (1.5,-.25) circle (3pt);
 \draw[ thick] (1.5,-.25) circle (3pt);
 \filldraw[green] (1.5+4,-.25) circle (3pt);
 \draw[ thick] (1.5+4,-.25) circle (3pt);
 \filldraw[green] (1.5+10,-.25) circle (3pt);
 \draw[thick] (1.5+10,-.25) circle (3pt);
 \filldraw[green] (1.5+14,-.25) circle (3pt);
 \draw[thick] (1.5+14,-.25) circle (3pt);
 \filldraw[green] (1.5+7,-1) circle (3pt);
 \draw[thick] (1.5+7,-1) circle (3pt);
 \node[scale = 1.5] at (8.5,0) { $\textbf{\dots}$};
 \node[scale = .5] at (.5,0) { $\textbf{\dots}$};
 \node[scale = .5] at (2.5,0) { $\textbf{\dots}$};
 \node[scale = .5] at (.5+4,0) { $\textbf{\dots}$};
 \node[scale = .5] at (2.5+4,0) { $\textbf{\dots}$};
 \node[scale = .5] at (.5+14,0) { $\textbf{\dots}$};
 \node[scale = .5] at (2.5+14,0) { $\textbf{\dots}$};
 \node[scale = .5] at (.5-2,0) { $\textbf{\dots}$};
 \node[scale = .5] at (2.5-2,0) { $\textbf{\dots}$};
 \draw [decorate,decoration={brace,amplitude=2pt,mirror,raise=2pt}]
 (1,.1) -- (0,.1) node[midway,yshift=9pt,scale=.5]{n};
 \end{tikzpicture}}\\
 \ket{R^{(k)}_2(\sigma_\gamma)} \;\;&\equiv\frac{1}{q^{n/2+1}} \;\; \scalebox{0.18}{\begin{tikzpicture}[baseline=-6ex]
 \tblockshortflip{0}{0}{0};
 \tblockshortflip{0}{12}{0};
 \tblockshortflip{0}{24+18}{0};
 \tblockshortflip{0}{36+18}{0};
 \wickU{4}{60}{7};
 \wickU{0}{64}{9};
 \wickU{6}{16}{5};
 \wickU{10}{12}{3};
 \wickU{48}{58}{5};
 \wickU{52}{54}{3};
 \draw[very thick] (18,-1.5) -- (18,{-.25*5-1.5}) -- (24, {-.25*5-1.5});
 \draw[very thick] (22,-1.5) -- (22,{-.25*3-1.5}) -- (24, {-.25*3-1.5});
 \draw[very thick] (46,-1.5) -- (46,{-.25*5-1.5}) -- (40, {-.25*5-1.5});
 \draw[very thick] (42,-1.5) -- (42,{-.25*3-1.5}) -- (40, {-.25*3-1.5});
 \draw[very thick] (-1,{\rt}) -- (-1,{-.25*11-1.5}) -- (10+36+18+1, {-.25*11-1.5}) -- (10+36+18+1, {\rt});
 \node[scale = 3] at (32,-0.5) { $\textbf{\dots}$};
 \filldraw[purple] (11,{\rt}) circle (6pt);
 \draw[thick] (11,{\rt}) circle (6pt);
 \filldraw[purple] (23,{\rt}) circle (6pt);
 \draw[thick] (23,{\rt}) circle (6pt);
 \filldraw[purple] (53,{\rt}) circle (6pt);
 \draw[thick] (53,{\rt}) circle (6pt);
 \filldraw[purple] (65,{\rt}) circle (6pt);
 \draw[thick] (65,{\rt}) circle (6pt);
 \end{tikzpicture}}.
\end{aligned}
\end{align}
The four possible OTOCs are thus
\begin{align}
\begin{aligned}
 C_{\alpha\gamma}^{(k),(11)}(x,t)&=\bra{L_1^{(k)}(\sigma_\alpha)} (T^{(k)}_{n})^{m} \ket{R_1^{(k)}(\sigma_\gamma)}\\
 C_{\alpha\gamma}^{(k),(21)}(x,t)&=\bra{L_2^{(k)}(\sigma_\alpha)} (T^{(k)}_{n})^{m} \ket{R_1^{(k)}(\sigma_\gamma)}\\ C_{\alpha\gamma}^{(k),(12)}(x,t)&=\bra{L_1^{(k)}(\sigma_\alpha)} (T^{(k)}_{n})^{m} \ket{R_2^{(k)}(\sigma_\gamma)}\\ C_{\alpha\gamma}^{(k),(22)}(x,t)&=\bra{L_2^{(k)}(\sigma_\alpha)} (T^{(k)}_{n})^{m} \ket{R_2^{(k)}(\sigma_\gamma)}
\end{aligned}.
\end{align}
The simplest setup is $C_{\alpha\gamma}(x,t)=\bra{L_1^{(k)}(\sigma_\alpha)} (T^{(k)}_{n})^{m} \ket{R_1^{(k)}(\sigma_\gamma)}$, the second possibility is $C_{\alpha\gamma}(x,t)=\bra{L_2^{(k)}(\sigma_\alpha)} (T^{(k)}_{n})^{m} \ket{R_1^{(k)}(\sigma_\gamma)}$, which is the same as $C_{\alpha\gamma}(x,t)=\bra{L_1^{(k)}(\sigma_\alpha)} (T^{(k)}_{n})^{m} \ket{R_2^{(k)}(\sigma_\gamma)}$ up to the following redefinition of $U$ and $U^\dagger$
\begin{align}
 \scalebox{.5}{\begin{tikzpicture}
 \U{0}{0}{0} 
 \node[scale = 2] at (3,.5) {$\rightarrow$};
 \U{180}{6}{1}
 \Ud{0}{9}{0} 
 \node[scale = 2] at (12,.5) {$\rightarrow$};
 \Ud{180}{15}{1}
 \end{tikzpicture}}
\end{align}
The third case with $\bra{L_2^{(k)}(\sigma_\alpha)} (T^{(k)}_{n})^{m} \ket{R_2^{(k)}(\sigma_\gamma)}$ is the most difficult.
Far along the light cone, 
\begin{align}
\lim_{m\rightarrow \infty}\left(T_n^{(k)}\right)^m = \frac{1}{q^{{kn}}} \sum_{n-chains}\ket{\tau_1 \geq \dots \geq \tau_n}\bra{\tau_1 \geq \dots \geq \tau_n}.
\end{align}

\subsubsection{Case I}

We start with the simplest set-up with $\bra{L_1^{(k)}(\sigma_\alpha)} (T^{(k)}_{n})^{m} \ket{R_1^{(k)}(\sigma_\gamma)}$. 
Consider $\tau_n$. In order for the projection to be non-zero on the left boundary, this must be a noncrossing permutation with only even length cycles, a set we denote $NC_e$. Even length cycles are necessary to form a product of an even number of Pauli matrices, which is the identity operator and hence has trace $q$. The product of an odd number of Pauli operators is traceless. Thus, 
\begin{align}
\label{eq:L1T}
\begin{aligned}
 \bra{L_1^{(k)}(\sigma_\alpha)}\sum_{n-chains \in NC}\ket{\tau_1 \geq \dots \geq \tau_n}\bra{\tau_1 \geq \dots \geq \tau_n}
 = q^{\sum_i C(\tau_i)}\sum_{n-chains \in NC_e}\bra{\tau_1 \geq \dots \geq \tau_n}.
\end{aligned}
\end{align}
All permutations consisting of only even length cycles must include at least two cycles involving nearest neighbors. This has a significant effect because, all vectors of this type are orthogonal to $\ket{R_1^{(k)}(\sigma_\gamma)}$ as the inner product is proportional to the trace of $\sigma_{\gamma}$. Therefore, all OTOCs of this parity are trivial at large $m$.

\subsubsection{Case II}
We now look at the second case, $\bra{L_1^{(k)}(\sigma_\alpha)} (T^{(k)}_{n})^{m} \ket{R_2^{(k)}(\sigma_\gamma)}$. The left boundary condition again requires that there be only even cycles in $\tau_{n}$. It is evident from \eqref{eq:L1T} that,
at large $q$, the permutations consisting of $k/2$ two-cycles dominate the projector when contracted with the left boundary. This fixes all $\tau_i$ to be the same. Using Kreweras's formula \eqref{eq:Kre}, there are
\begin{align}
 \frac{k!}{(k/2+1)!(k/2)!} = C_{k/2}\label{eq:2cyc}
\end{align}
of these, where $C_i$ is the $i^{th}$ Catalan number.

It is instructive to treat the $k=4$ case example explicitly. There are two elements in $NC_4$ with only two-cycles, $[1,2][3,4]$ and $[1,4][2,3]$. These give equal contributions because they are related by cyclicity. Therefore, we only need to consider the $\tau_n=[1,2][3,4] = \tau_i$ term, which is proportional to 
\begin{align}
 &\;\;\scalebox{0.92}{\begin{tikzpicture}[baseline=0]
 \node[] at (10.5,0) {$\dots$};
 \node[] at (8.5,0) {$\dots$};
 \node[] at (12.5,0) {$\dots$};
 \node[] at (14.5,0) {$\dots$};
 \node[] at (6.5,0) {$\dots$};
 \node[] at (4.5,0) {$\dots$};
 \node[] at (2.5,0) {$\dots$};
 \node[] at (0.5,0) {$\dots$};
 \wickflip{1}{2}{1}{0}{0}
 \wickflip{0}{3}{2}{0}{0}
 \wickflip{1+4}{2+4}{1}{0}{0}
 \wickflip{0+4}{3+4}{2}{0}{0}
 \wickflip{1+8}{2+8}{1}{0}{0}
 \wickflip{0+8}{3+8}{2}{0}{0}
 \wickflip{1+12}{2+12}{1}{0}{0}
 \wickflip{0+12}{3+12}{2}{0}{0}
 \filldraw[orange] (1.5,.25) circle (3pt);
 \draw[ thick] (1.5,.25) circle (3pt);
 \filldraw[orange] (1.5+4,.25) circle (3pt);
 \draw[ thick] (1.5+4,.25) circle (3pt);
 \filldraw[orange] (1.5+8,.25) circle (3pt);
 \draw[ thick] (1.5+8,.25) circle (3pt);
 \filldraw[orange] (1.5+12,.25) circle (3pt);
 \draw[ thick] (1.5+12,.25) circle (3pt);
 \wick{0}{7}{4}
 \wick{1}{6}{3}
 \wick{2}{5}{2}
 \wick{3}{4}{1}
 \wick{0+8}{7+8}{4}
 \wick{1+8}{6+8}{3}
 \wick{2+8}{5+8}{2}
 \wick{3+8}{4+8}{1}
 \end{tikzpicture}}\nonumber\\
 &\scalebox{0.3}{
 \begin{tikzpicture}
 \tblockshortflip{0}{0}{0}
 \tblockshortflip{0}{12}{0}
 \tblockshortflip{0}{24}{0}
 \tblockshortflip{0}{36}{0}
 \wickU{4}{42}{7};
 \wickU{0}{46}{9};
 \wickU{6}{16}{5};
 \wickU{10}{12}{3};
 \wickU{18}{28}{5};
 \wickU{22}{24}{3};
 \wickU{18+12}{28+12}{5};
 \wickU{22+12}{24+12}{3};
 \draw[very thick] (-1,{\rt}) -- (-1,{-.25*11-1.5}) -- (47, {-.25*11-1.5}) -- (47, {\rt});
 \draw[thick] (11,{\rt}) circle (3pt);
 \filldraw[purple] (11,{\rt}) circle (3pt);
 \draw[thick] (23,{\rt}) circle (3pt);
 \filldraw[purple] (23,{\rt}) circle (3pt);
 \draw[thick] (35,{\rt}) circle (3pt);
 \filldraw[purple] (35,{\rt}) circle (3pt);
 \draw[thick] (35+12,{\rt}) circle (3pt);
 \filldraw[purple] (35+12,{\rt}) circle (3pt);
 \wickflipU{0}{22}{9}
 \wickflipU{4}{18}{7}
 \wickflipU{6}{16}{5}
 \wickflipU{10}{12}{3}
 \wickflipU{0+24}{22+24}{9}
 \wickflipU{4+24}{18+24}{7}
 \wickflipU{6+24}{16+24}{5}
 \wickflipU{10+24}{12+24}{3}
 \end{tikzpicture}}\nonumber\\
 &\propto \mathcal{N}_1^n(\sigma_\gamma)_{ab} \;\mathcal{N}_2^n(\sigma_\gamma \otimes\sigma_\gamma)_{ab,cd} \; \mathcal{N}_1^n(\sigma_\gamma)_{dc}\label{eq:12 34}
\end{align}
where the sums over repeated indices is implied.
Here, we define quantum channels generalizing ${\mathcal{N}}_1 \equiv \mathcal{M}_+$
\begin{align}
\begin{aligned}
&\tilde{\mathcal{N}}_m(\sigma_1\otimes \sigma_2\otimes \cdots \otimes \sigma_m) 
 \scalebox{0.5}{
 \begin{tikzpicture}[baseline=-6ex]
 \node[scale = 3] at (-3,-1) {$\equiv \frac{1}{q}$};
 \Ud{-45}{{\rt+2}}{{\rt}}\U{-135}{{0}}{{0}}
 \draw[very thick] (0,0) -- (0,1) -- (17,1) -- (17,0);
 \filldraw[purple] (1,{\rt}) circle (5pt);
 \draw[very thick] (1,{\rt}) circle (5pt);
 \node[scale = 1] at (1,-.2) {$\sigma_1$};
 \draw[very thick] (0,{-1-\rt-1}) -- (0,{-1.5-\rt-1})--(2,{-1.5-\rt-1}) --(2,{-1-\rt-1});
 \Ud{-45}{{\rt+5+2}}{{\rt}}\U{-135}{{5}}{{0}}
 \draw[very thick] (2,{\rt+1}) -- (5,{\rt+1});
 \draw[very thick] (2+10,{\rt+1}) -- (5+10,{\rt+1});
 \draw[very thick] (0+5,{-1-\rt-1}) -- (0+5,{-1.5-\rt-1})--(2+5,{-1.5-\rt-1}) --(2+5,{-1-\rt-1});
 \filldraw[purple] (1+5,{\rt}) circle (5pt);
 \draw[very thick] (1+5,{\rt}) circle (5pt);
 \Ud{-45}{{\rt+10+2}}{{\rt}}\U{-135}{{10}}{{0}}
 \draw[very thick] (0+10,{-1-\rt-1}) -- (0+10,{-1.5-\rt-1})--(2+10,{-1.5-\rt-1}) --(2+10,{-1-\rt-1});
 \filldraw[purple] (1+10,{\rt}) circle (5pt);
 \draw[very thick] (1+10,{\rt}) circle (5pt);
 \node[scale = 1] at (1+10,-.2) {$\sigma_{m-1}$};
 \Ud{-45}{{\rt+5+10+2}}{{\rt}}\U{-135}{{5+10}}{{0}}
 \draw[very thick] (2+10,{\rt+1}) -- (5+10,{\rt+1});
 \draw[very thick] (2+5,{\rt+1}) -- (3+5,{\rt+1});
 \draw[very thick] (4+5,{\rt+1}) -- (5+5,{\rt+1});
 \draw[very thick] (0+15,{-1-\rt-1}) -- (0+15,{-1.5-\rt-1})--(2+15,{-1.5-\rt-1}) --(2+15,{-1-\rt-1});
 \filldraw[purple] (1+5+10,{\rt}) circle (5pt);
 \draw[very thick] (1+5+10,{\rt}) circle (5pt);
 \node[scale = 1.5] at (8.5,-1) {$\cdots$};
 \end{tikzpicture}}\\
 &
 \scalebox{.5}{
 \begin{tikzpicture}
 \U{180}{0}{0}
 \Ud{0}{1}{{-1}}
 \U{180}{4}{0}
 \Ud{0}{6}{{-1}}
 \U{180}{9}{0}
 \Ud{0}{10}{{-1}}
 \Ud{180}{0}{-3}
 \U{0}{1}{{-4}}
 \Ud{180}{4}{-3}
 \U{0}{6}{{-4}}
 \Ud{180}{9}{-3}
 \U{0}{10}{{-4}}
 \draw[very thick] (-1.25,-1.25) -- (-1.25,-2.75) ;
 \draw[very thick] (.25,-1.25) -- (.25,-2.75) ;
 \draw[very thick] (9.75,-1.25) -- (9.75,-2.75) ;
 \draw[very thick] (11.25,-1.25) -- (11.25,-2.75) ;
 \draw[very thick] (2.25,-1.25) -- (2.75,-1.25) ;
 \draw[very thick] (2.25,-2.75) -- (2.75,-2.75) ;
 \draw[very thick] (7.25,-1.25) -- (7.75,-1.25) ;
 \draw[very thick] (7.25,-2.75) -- (7.75,-2.75) ;
 \draw[very thick] (5.75,-2.75) -- (5.75,-2.25) -- (9.25,-2.25) -- (9.25,-2.75) ;
 \draw[very thick] (.75,-2.75) -- (.75,-2.25) -- (4.25,-2.25) -- (4.25,-2.75) ;
 \draw[very thick] (5.75,-1.25) -- (5.75,-1.75) -- (9.25,-1.75) -- (9.25,-1.25) ;
 \draw[very thick] (.75,-1.25) -- (.75,-1.75) -- (4.25,-1.75) -- (4.25,-1.25) ;
 \draw[very thick] (.25,.25) -- (.75,.25) ;
 \draw[very thick] (.25,-4.25) -- (.75,-4.25) ;
 \draw[very thick] (4.25,.25) -- (4.5,.25) ;
 \draw[very thick] (4.25,-4.25) -- (4.5,-4.25) ;
 \draw[very thick] (5.5,.25) -- (5.75,.25) ;
 \draw[very thick] (5.5,-4.25) -- (5.75,-4.25) ;
 \draw[very thick] (9.25,.25) -- (9.75,.25) ;
 \draw[very thick] (9.25,-4.25) -- (9.75,-4.25) ;
 \node[scale = 1.5] at (5,-2) {$\dots$};
 \node[scale = 1.5] at (5,.25) {$\dots$};
 \node[scale = 1.5] at (5.,-4.25) {$\dots$};
 \draw [decorate,decoration={brace,amplitude=6pt}]
 (9.5,-5) -- (0.5,-5) node[midway,yshift=-2.5em,scale=2]{$\frac{k-2}{2}$};
 \node[scale = 3] at (13.5, -2) {$=$};
 \node[scale = 3] at (-3, -2) {$= \frac{1}{q}$};
 \filldraw[yellow] (16,-1) rectangle (24,0); 
 \draw[very thick] (16,-1) rectangle (24,0); 
 \filldraw[yellow] (16,-4) rectangle (24,-3); 
 \draw[very thick] (16,-4) rectangle (24,-3); 
 \draw[very thick] (16.5,-1) -- (16.5,-3);
 \draw[very thick] (16.5,0) -- (16.5,.5) -- (15.5,.5) --(15.5,-4.5) --(16.5,-4.5) -- (16.5,-4);
 \draw[very thick] (17,0) -- (17,.5);
 \draw[very thick] (18,0) -- (18,.5);
 \draw[very thick] (17,-4) -- (17,-4.5);
 \draw[very thick] (18,-4) -- (18,-4.5);
 \draw[very thick] (17, -1) -- (17, -1.5) -- (18, -1.5) -- (18, -1);
 \draw[very thick] (17, -3) -- (17, -2.5) -- (18, -2.5) -- (18, -3);
 \draw[very thick] (21.5, -1) -- (21.5, -1.5) -- (22.5, -1.5) -- (22.5, -1);
 \draw[very thick] (21.5,0) -- (21.5,.5);
 \draw[very thick] (21.5,-4) -- (21.5,-4.5);
 \draw[very thick] (22.5,0) -- (22.5,.5);
 \draw[very thick] (22.5,-4) -- (22.5,-4.5);
 \draw[very thick] (21.5, -3) -- (21.5, -2.5) -- (22.5, -2.5) -- (22.5, -3);
 \draw[very thick] (23,-1) -- (23,-3);
 \draw[very thick] (23.5,-1) -- (23.5,-3);
 \draw[very thick] (23.5,0) -- (23.5,.5);
 \draw[very thick] (23.5,-4) -- (23.5,-4.5);
 \draw[very thick] (23,0) -- (23,.5);
 \draw[very thick] (23,-4) -- (23,-4.5);
 \node[scale = 2] at (19.75,-2) {$\dots$};
 \node[scale = 2] at (19.75,.25) {$\dots$};
 \node[scale = 2] at (19.75,-4.25) {$\dots$};
 \node[scale = 2] at (19.75,-3.5) {$V^{\dagger}$};
 \node[scale = 2] at (19.75,-.5) {$V$};
 \filldraw[purple] (0.25,-2) circle (5pt);
 \draw[very thick] (0.25,-2) circle (5pt);
 \filldraw[purple] (9.75,-2) circle (5pt);
 \draw[very thick] (9.75,-2) circle (5pt);
 \filldraw[purple] (7.5,-2.25) circle (5pt);
 \draw[very thick] (7.5,-2.25) circle (5pt); 
 \filldraw[purple] (7.5,-1.75) circle (5pt);
 \draw[very thick] (7.5,-1.75) circle (5pt); 
 \filldraw[purple] (2.5,-2.25) circle (5pt);
 \draw[very thick] (2.5,-2.25) circle (5pt); 
 \filldraw[purple] (2.5,-1.75) circle (5pt);
 \draw[very thick] (2.5,-1.75) circle (5pt); 
 \filldraw[purple] (16.5,-2) circle (5pt);
 \draw[very thick] (16.5,-2) circle (5pt); 
 \filldraw[purple] (23,-2) circle (5pt);
 \draw[very thick] (23,-2) circle (5pt); \filldraw[purple] (22,-1.5) circle (5pt);
 \draw[very thick] (22,-1.5) circle (5pt); \filldraw[purple] (22,-2.5) circle (5pt);
 \draw[very thick] (22,-2.5) circle (5pt); \filldraw[purple] (17.5,-1.5) circle (5pt);
 \draw[very thick] (17.5,-1.5) circle (5pt); \filldraw[purple] (17.5,-2.5) circle (5pt);
 \draw[very thick] (17.5,-2.5) circle (5pt); 
 \end{tikzpicture}}
 \\
 &= \mathcal{N}_m(\sigma_\gamma \otimes\mathcal{P}_{\sigma_\gamma}^{{\otimes}^\frac{m-2}{2}} \otimes\sigma_\gamma)
 \end{aligned},
\end{align}
where $\mathcal{P}_{\sigma_\gamma}\equiv \ket{\sigma_\gamma} \bra{\sigma_\gamma}$ and $\ket{\mathcal{O}}$ is the state obtained by viewing at the operator ${\mathcal{O}}$ as a vector on $\mathbb{C}^q\otimes \mathbb{C}^q$, the so-called Choi–Jamiolkowski isomorphism \cite{choi1975completely,jamiolkowski1972linear}. 
We have furthermore defined $V$ as
\begin{align}\scalebox{.5}{
\begin{tikzpicture}
 \filldraw[yellow] (-3,-1) rectangle (5,0); 
 \draw[very thick] (-3,-1) rectangle (5,0); 
 \draw[very thick] (-2.5,0)--(-2.5,0.5);
 \draw[very thick] (-1.5,0)--(-1.5,0.5);
 \draw[very thick] (-.5,0)--(-.5,0.5);
 \draw[very thick] (-2.5,-1)--(-2.5,-1.5);
 \draw[very thick] (-1.5,-1)--(-1.5,-1.5);
 \draw[very thick] (-.5,-1)--(-.5,-1.5);
 \draw[very thick] (2.5,0)--(2.5,0.5);
 \draw[very thick] (3.5,0)--(3.5,0.5);
 \draw[very thick] (4.5,0)--(4.5,0.5);
 \draw[very thick] (2.5,-1)--(2.5,-1.5);
 \draw[very thick] (3.5,-1)--(3.5,-1.5);
 \draw[very thick] (4.5,-1)--(4.5,-1.5);
 \node[scale = 2] at (1,-.5) {$V$};
 \node[scale = 2] at (1,-1.25) {$\dots$};
 \node[scale = 2] at (1,.25) {$\dots$};
 \U{180}{10}{0}
 \Ud{0}{11}{{-1}}
 \U{180}{14}{0}
 \Ud{0}{16}{{-1}}
 \U{180}{19}{0}
 \Ud{0}{20}{{-1}}
 \draw[very thick] (8.75,-1.25) -- (8.25,-1.25) -- (8.25 , 0.25) ; 
 \draw[very thick] (10.25,.25) -- (10.75,.25); 
 \draw[very thick] (12.25,-1.25) -- (12.75,-1.25); 
 \draw[very thick] (17.25,-1.25) -- (17.75,-1.25); 
 \draw[very thick] (19.25,.25) -- (19.75,.25); 
 \draw[very thick] (15.5,.25) -- (15.75,.25); 
 \draw[very thick] (14.5,.25) -- (14.25,.25); 
 \node[scale = 2] at (15,-.5) {$\dots$};
 \node[scale = 2] at (6.5,-.5) {$\equiv$};
\end{tikzpicture}}
\end{align}
The unitarity of $V$ ($V V^{\dagger} = V^{\dagger} V = \mathbbm{1}$) follows from applying the unitarity and dual unitarity conditions of $U$ $m$ times. $\mathcal{N}_m$ can then be seen to be a unital quantum channels because it is the composition of a unitary and a partial trace.

From the $k=4$ example, we see that we will need the generalized channels in the expressions of the higher $k$ OTOCs. In fact, these are the only components we will need. 
The argument proceeds by representing the overall trace of the right boundary condition as a circle, which is the natural choice given the symmetry. Blue and red segments represents the $U$ and $U^\dagger$ operators. The purple dots denote the operators on the right boundary condition ($\sigma_\gamma$) and orange dots denote the ``dual lattice'' points where the corresponding operators would be on the left boundary condition ($\sigma_\alpha$). We can then represent the $[1,2][3,4]$ permutation in \eqref{eq:12 34} as the diagram
\begin{align}
 \begin{tikzpicture}[scale=1.5,baseline=-4]
 \centerarc[\darkblue,very thick](0,0)(0:45:20pt)
 \centerarc[\darkred,very thick](0,0)(45:90:20pt)
 \centerarc[\darkblue,very thick](0,0)(90:135:20pt)
 \centerarc[\darkred,very thick](0,0)(135:180:20pt)
 \centerarc[\darkblue,very thick](0,0)(180:225:20pt)
 \centerarc[\darkred,very thick](0,0)(225:270:20pt)
 \centerarc[\darkblue,very thick](0,0)(270:315:20pt)
 \centerarc[\darkred,very thick](0,0)(315:360:20pt)
 \filldraw[purple] (.7,0) circle (3pt);
 \filldraw[purple] (0,.7) circle (3pt);
 \filldraw[purple] (-.7,0) circle (3pt);
 \filldraw[purple] (0,-.7) circle (3pt);
 \filldraw[orange] ({.7*\rt},{.7*\rt}) circle (3pt);
 \filldraw[orange] ({-.7*\rt},{.7*\rt}) circle (3pt);
 \filldraw[orange] ({.7*\rt},{-.7*\rt}) circle (3pt);
 \filldraw[orange] ({-.7*\rt},{-.7*\rt}) circle (3pt);
 \draw[very thick] (.7,0) circle (3pt);
 \draw[very thick] (0,.7) circle (3pt);
 \draw[very thick] (-.7,0) circle (3pt);
 \draw[very thick] (0,-.7) circle (3pt);
 \draw[very thick] ({.7*\rt},{.7*\rt}) circle (3pt);
 \draw[very thick] ({-.7*\rt},{.7*\rt}) circle (3pt);
 \draw[very thick] ({.7*\rt},{-.7*\rt}) circle (3pt);
 \draw[very thick] ({-.7*\rt},{-.7*\rt}) circle (3pt);
 \draw[very thick] ({cos(75)*20pt},{sin(75)*20pt}) .. controls ({cos(75)*20pt-5},{sin(75)*20pt-5}) and ({cos(60+45)*20pt+5},{sin(60+45)*20pt-5}) .. ({cos(60+45)*20pt},{sin(60+45)*20pt});
 \draw[very thick] ({cos(60)*20pt},{sin(60)*20pt}) .. controls ({cos(60)*20pt-5},{sin(60)*20pt-5}) and ({cos(120)*20pt+5},{sin(120)*20pt-5}) .. ({cos(120)*20pt},{sin(120)*20pt});
 \draw[very thick] ({cos(75+180)*20pt},{sin(75+180)*20pt}) .. controls ({cos(75+180)*20pt+5},{sin(75+180)*20pt+5}) and ({cos(60+45+180)*20pt-5},{sin(60+45+180)*20pt+5}) .. ({cos(60+45+180)*20pt},{sin(60+45+180)*20pt});
 \draw[very thick] ({cos(60+180)*20pt},{sin(60+180)*20pt}) .. controls ({cos(60+180)*20pt+5},{sin(60+180)*20pt+5}) and ({cos(120+180)*20pt-5},{sin(120+180)*20pt+5}) .. ({cos(120+180)*20pt},{sin(120+180)*20pt});
 \draw[very thick] ({cos(150)*20pt},{sin(150)*20pt}) .. controls ({cos(150)*20pt-5},{sin(150)*20pt-5}) and ({cos(30)*20pt+5},{sin(30)*20pt-5}) .. ({cos(30)*20pt},{sin(30)*20pt});
 \draw[very thick] ({cos(160)*20pt},{sin(160)*20pt}) .. controls ({cos(160)*20pt-5},{sin(160)*20pt-5}) and ({cos(20)*20pt+5},{sin(20)*20pt-5}) .. ({cos(20)*20pt},{sin(20)*20pt});
 \draw[very thick] ({cos(-150)*20pt},{sin(-150)*20pt}) .. controls ({cos(-150)*20pt-5},{sin(-150)*20pt+5}) and ({cos(-30)*20pt-5},{sin(-30)*20pt+5}) .. ({cos(-30)*20pt},{sin(-30)*20pt});
 \draw[very thick] ({cos(-160)*20pt},{sin(-160)*20pt}) .. controls ({cos(-160)*20pt-5},{sin(-160)*20pt+5}) and ({cos(-20)*20pt-5},{sin(-20)*20pt+5}) .. ({cos(-20)*20pt},{sin(-20)*20pt});
 \node[scale = .3] at (0,14.5pt) {$\vdots$};
 \node[scale = .3] at (0,-13.5pt) {$\vdots$};
 \node[scale = .3] at (0,5pt) {$\vdots$};
 \node[scale = .3] at (0,-4pt) {$\vdots$};
 \end{tikzpicture}\end{align}
There are $n$ lines connecting red and blue segments (the ``$\dots$'' between the solid lines represents $n-2$ hidden lines), representing the eigenvector of the transfer matrix which contracts with the right boundary condition. Other contractions at arbitrary $k$ can be represented similarly.

Viewing these as permutations $\tau_i$ on Pauli operators on the left boundary condition 
amounts to joining the two lines adjacent to an orange dot across the orange dot, giving the graphical representation of $\tau_n$. This is repeated $n-1$ more times to get $\tau_{n-1}$ through $\tau_1$. In the above example, $\tau_i = [1,2][3,4]$ corresponding to the two sets of loops in the diagram
\begin{align}
 \begin{tikzpicture}[scale=1.5,baseline=-4]
 \centerarc[black,very thick](0,0)(30:60:20pt)
 \centerarc[black,very thick](0,0)(30+90:60+90:20pt)
 \centerarc[black,very thick](0,0)(30+180:60+180:20pt)
 \centerarc[black,very thick](0,0)(30+270:60+270:20pt)
 \filldraw[purple] (.7,0) circle (3pt);
 \filldraw[purple] (0,.7) circle (3pt);
 \filldraw[purple] (-.7,0) circle (3pt);
 \filldraw[purple] (0,-.7) circle (3pt);
 \filldraw[orange] ({.7*\rt},{.7*\rt}) circle (3pt);
 \filldraw[orange] ({-.7*\rt},{.7*\rt}) circle (3pt);
 \filldraw[orange] ({.7*\rt},{-.7*\rt}) circle (3pt);
 \filldraw[orange] ({-.7*\rt},{-.7*\rt}) circle (3pt);
 \draw[very thick] (.7,0) circle (3pt);
 \draw[very thick] (0,.7) circle (3pt);
 \draw[very thick] (-.7,0) circle (3pt);
 \draw[very thick] (0,-.7) circle (3pt);
 \draw[very thick] ({.7*\rt},{.7*\rt}) circle (3pt);
 \draw[very thick] ({-.7*\rt},{.7*\rt}) circle (3pt);
 \draw[very thick] ({.7*\rt},{-.7*\rt}) circle (3pt);
 \draw[very thick] ({-.7*\rt},{-.7*\rt}) circle (3pt);
 \draw[very thick] ({cos(75)*20pt},{sin(75)*20pt}) .. controls ({cos(75)*20pt-5},{sin(75)*20pt-5}) and ({cos(60+45)*20pt+5},{sin(60+45)*20pt-5}) .. ({cos(60+45)*20pt},{sin(60+45)*20pt});
 \draw[very thick] ({cos(60)*20pt},{sin(60)*20pt}) .. controls ({cos(60)*20pt-5},{sin(60)*20pt-5}) and ({cos(120)*20pt+5},{sin(120)*20pt-5}) .. ({cos(120)*20pt},{sin(120)*20pt});
 \draw[very thick] ({cos(75+180)*20pt},{sin(75+180)*20pt}) .. controls ({cos(75+180)*20pt+5},{sin(75+180)*20pt+5}) and ({cos(60+45+180)*20pt-5},{sin(60+45+180)*20pt+5}) .. ({cos(60+45+180)*20pt},{sin(60+45+180)*20pt});
 \draw[very thick] ({cos(60+180)*20pt},{sin(60+180)*20pt}) .. controls ({cos(60+180)*20pt+5},{sin(60+180)*20pt+5}) and ({cos(120+180)*20pt-5},{sin(120+180)*20pt+5}) .. ({cos(120+180)*20pt},{sin(120+180)*20pt});
 \draw[very thick] ({cos(150)*20pt},{sin(150)*20pt}) .. controls ({cos(150)*20pt-5},{sin(150)*20pt-5}) and ({cos(30)*20pt+5},{sin(30)*20pt-5}) .. ({cos(30)*20pt},{sin(30)*20pt});
 \draw[very thick] ({cos(160)*20pt},{sin(160)*20pt}) .. controls ({cos(160)*20pt-5},{sin(160)*20pt-5}) and ({cos(20)*20pt+5},{sin(20)*20pt-5}) .. ({cos(20)*20pt},{sin(20)*20pt});
 \draw[very thick] ({cos(-150)*20pt},{sin(-150)*20pt}) .. controls ({cos(-150)*20pt-5},{sin(-150)*20pt+5}) and ({cos(-30)*20pt-5},{sin(-30)*20pt+5}) .. ({cos(-30)*20pt},{sin(-30)*20pt});
 \draw[very thick] ({cos(-160)*20pt},{sin(-160)*20pt}) .. controls ({cos(-160)*20pt-5},{sin(-160)*20pt+5}) and ({cos(-20)*20pt-5},{sin(-20)*20pt+5}) .. ({cos(-20)*20pt},{sin(-20)*20pt});
 \node[scale = .3] at (0,14.5pt) {$\vdots$};
 \node[scale = .3] at (0,-13.5pt) {$\vdots$};
 \node[scale = .3] at (0,5pt) {$\vdots$};
 \node[scale = .3] at (0,-4pt) {$\vdots$};
 \draw[very thick] ({cos(75)*20pt},{sin(75)*20pt}) .. controls ({cos(75)*20pt+10},{sin(75)*20pt+5}) and ({cos(15)*20pt+5},{sin(15)*20pt+10}) .. ({cos(15)*20pt},{sin(15)*20pt});
 \draw[very thick] ({cos(105)*20pt},{sin(105)*20pt}) .. controls ({cos(105)*20pt-10},{sin(105)*20pt+5}) and ({cos(165)*20pt-5},{sin(165)*20pt+10}) .. ({cos(165)*20pt},{sin(165)*20pt});
 \draw[very thick] ({cos(-75)*20pt},{sin(-75)*20pt}) .. controls ({cos(-75)*20pt+10},{sin(-75)*20pt-5}) and ({cos(-20)*20pt+5},{sin(-20)*20pt-10}) .. ({cos(-20)*20pt},{sin(-20)*20pt});
 \draw[very thick] ({cos(-105)*20pt},{sin(-105)*20pt}) .. controls ({cos(-105)*20pt-10},{sin(-105)*20pt-5}) and ({cos(-165)*20pt-5},{sin(-165)*20pt-10}) .. ({cos(-165)*20pt},{sin(-165)*20pt});
 \end{tikzpicture}
 &=\;\scalebox{0.8}{\begin{tikzpicture}[baseline=0]
 \wickflip{1}{2}{1}{0}{0}
 \wickflip{0}{3}{2}{0}{0}
 \wickflip{1+4}{2+4}{1}{0}{0}
 \wickflip{0+4}{3+4}{2}{0}{0}
 \wickflip{1+8}{2+8}{1}{0}{0}
 \wickflip{0+8}{3+8}{2}{0}{0}
 \wickflip{1+12}{2+12}{1}{0}{0}
 \wickflip{0+12}{3+12}{2}{0}{0}
 \filldraw[orange] (1.5,.25) circle (3pt);
 \draw[ thick] (1.5,.25) circle (3pt);
 \filldraw[orange] (1.5+4,.25) circle (3pt);
 \draw[ thick] (1.5+4,.25) circle (3pt);
 \filldraw[orange] (1.5+8,.25) circle (3pt);
 \draw[ thick] (1.5+8,.25) circle (3pt);
 \filldraw[orange] (1.5+12,.25) circle (3pt);
 \draw[ thick] (1.5+12,.25) circle (3pt);
 \wick{0}{7}{4}
 \wick{1}{6}{3}
 \wick{2}{5}{2}
 \wick{3}{4}{1}
 \wick{0+8}{7+8}{4}
 \wick{1+8}{6+8}{3}
 \wick{2+8}{5+8}{2}
 \wick{3+8}{4+8}{1}
 \node[] at (10.5,0) {$\dots$};
 \node[] at (8.5,0) {$\dots$};
 \node[] at (12.5,0) {$\dots$};
 \node[] at (14.5,0) {$\dots$};
 \node[] at (6.5,0) {$\dots$};
 \node[] at (4.5,0) {$\dots$};
 \node[] at (2.5,0) {$\dots$};
 \node[] at (0.5,0) {$\dots$};
 \end{tikzpicture}}
\end{align}
On the other hand, to see the action of quantum channels on $\sigma_\gamma$ on the right, we note that the right boundary condition connects the lines adjacent to the purple dots with ``rainbows'' on the outside of the purple dots, which gives the graphical representation of the Kreweras complements of $\tau_1$ through $\tau_n$. The Kreweras complement arises because we are looking at the action of the contraction lines of a permutation as a permutation on the dual lattice points, which is the maximal complement permutation. In the $[1,2][3,4]$ example above, we have that $K([1,2][3,4])=[1][2,4][3]$. This corresponds to $\mathcal{N}^n_2$ composed with two $\mathcal{N}^n_1$, which are given by the three sets of black loops in the diagram. These are composed as $\bra{\mathcal{N}^n_1(\sigma_\gamma)}\mathcal{N}_2^n(\sigma_\gamma \otimes\sigma_\gamma)\ket{\mathcal{N}^n_1(\sigma_\gamma)}$ according to the placement of the cycles
\begin{align}
 \begin{tikzpicture}[scale=1.5,baseline=-4]
 \centerarc[\darkblue,very thick](0,0)(0:45:20pt)
 \centerarc[\darkred,very thick](0,0)(45:90:20pt)
 \centerarc[\darkblue,very thick](0,0)(90:135:20pt)
 \centerarc[\darkred,very thick](0,0)(135:180:20pt)
 \centerarc[\darkblue,very thick](0,0)(180:225:20pt)
 \centerarc[\darkred,very thick](0,0)(225:270:20pt)
 \centerarc[\darkblue,very thick](0,0)(270:315:20pt)
 \centerarc[\darkred,very thick](0,0)(315:360:20pt)
 \filldraw[purple] (.7,0) circle (3pt);
 \filldraw[purple] (0,.7) circle (3pt);
 \filldraw[purple] (-.7,0) circle (3pt);
 \filldraw[purple] (0,-.7) circle (3pt);
 \filldraw[orange] ({.7*\rt},{.7*\rt}) circle (3pt);
 \filldraw[orange] ({-.7*\rt},{.7*\rt}) circle (3pt);
 \filldraw[orange] ({.7*\rt},{-.7*\rt}) circle (3pt);
 \filldraw[orange] ({-.7*\rt},{-.7*\rt}) circle (3pt);
 \draw[very thick] (.7,0) circle (3pt);
 \draw[very thick] (0,.7) circle (3pt);
 \draw[very thick] (-.7,0) circle (3pt);
 \draw[very thick] (0,-.7) circle (3pt);
 \draw[very thick] ({.7*\rt},{.7*\rt}) circle (3pt);
 \draw[very thick] ({-.7*\rt},{.7*\rt}) circle (3pt);
 \draw[very thick] ({.7*\rt},{-.7*\rt}) circle (3pt);
 \draw[very thick] ({-.7*\rt},{-.7*\rt}) circle (3pt);
 \draw[very thick] ({cos(75)*20pt},{sin(75)*20pt}) .. controls ({cos(75)*20pt-5},{sin(75)*20pt-5}) and ({cos(60+45)*20pt+5},{sin(60+45)*20pt-5}) .. ({cos(60+45)*20pt},{sin(60+45)*20pt});
 \draw[very thick] ({cos(60)*20pt},{sin(60)*20pt}) .. controls ({cos(60)*20pt-5},{sin(60)*20pt-5}) and ({cos(120)*20pt+5},{sin(120)*20pt-5}) .. ({cos(120)*20pt},{sin(120)*20pt});
 \draw[very thick] ({cos(75+180)*20pt},{sin(75+180)*20pt}) .. controls ({cos(75+180)*20pt+5},{sin(75+180)*20pt+5}) and ({cos(60+45+180)*20pt-5},{sin(60+45+180)*20pt+5}) .. ({cos(60+45+180)*20pt},{sin(60+45+180)*20pt});
 \draw[very thick] ({cos(60+180)*20pt},{sin(60+180)*20pt}) .. controls ({cos(60+180)*20pt+5},{sin(60+180)*20pt+5}) and ({cos(120+180)*20pt-5},{sin(120+180)*20pt+5}) .. ({cos(120+180)*20pt},{sin(120+180)*20pt});
 \draw[very thick] ({cos(150)*20pt},{sin(150)*20pt}) .. controls ({cos(150)*20pt-5},{sin(150)*20pt-5}) and ({cos(30)*20pt+5},{sin(30)*20pt-5}) .. ({cos(30)*20pt},{sin(30)*20pt});
 \draw[very thick] ({cos(160)*20pt},{sin(160)*20pt}) .. controls ({cos(160)*20pt-5},{sin(160)*20pt-5}) and ({cos(20)*20pt+5},{sin(20)*20pt-5}) .. ({cos(20)*20pt},{sin(20)*20pt});
 \draw[very thick] ({cos(-150)*20pt},{sin(-150)*20pt}) .. controls ({cos(-150)*20pt-5},{sin(-150)*20pt+5}) and ({cos(-30)*20pt-5},{sin(-30)*20pt+5}) .. ({cos(-30)*20pt},{sin(-30)*20pt});
 \draw[very thick] ({cos(-160)*20pt},{sin(-160)*20pt}) .. controls ({cos(-160)*20pt-5},{sin(-160)*20pt+5}) and ({cos(-20)*20pt-5},{sin(-20)*20pt+5}) .. ({cos(-20)*20pt},{sin(-20)*20pt});
 \node[scale = .3] at (0,14.5pt) {$\vdots$};
 \node[scale = .3] at (0,-13.5pt) {$\vdots$};
 \node[scale = .3] at (0,5pt) {$\vdots$};
 \node[scale = .3] at (0,-4pt) {$\vdots$};
 \draw[very thick] ({cos(75)*20pt},{sin(75)*20pt}) .. controls ({cos(75)*20pt-3},{sin(75)*20pt+7}) and ({cos(60+45)*20pt+3},{sin(60+45)*20pt+7}) .. ({cos(60+45)*20pt},{sin(60+45)*20pt});
 \draw[very thick] ({cos(60)*20pt},{sin(60)*20pt}) .. controls ({cos(60)*20pt-5},{sin(60)*20pt+14}) and ({cos(120)*20pt+5},{sin(120)*20pt+14}) .. ({cos(120)*20pt},{sin(120)*20pt});
 \draw[very thick] ({cos(-75)*20pt},{sin(-75)*20pt}) .. controls ({cos(-75)*20pt+3},{sin(-75)*20pt-7}) and ({cos(-60-45)*20pt-3},{sin(-60-45)*20pt-7}) .. ({cos(-60+-45)*20pt},{sin(-60+-45)*20pt});
 \draw[very thick] ({cos(-60)*20pt},{sin(-60)*20pt}) .. controls ({cos(-60)*20pt+5},{sin(-60)*20pt-14}) and ({cos(-120)*20pt-5},{sin(-120)*20pt-14}) .. ({cos(-120)*20pt},{sin(-120)*20pt});
 \draw[very thick] ({cos(150)*20pt},{sin(150)*20pt}) .. controls ({cos(150)*20pt-14},{sin(150)*20pt-5}) and ({cos(-150)*20pt-14},{sin(-150)*20pt+5}) .. ({cos(-150)*20pt},{sin(-150)*20pt});
 \draw[very thick] ({cos(160)*20pt},{sin(160)*20pt}) .. controls ({cos(160)*20pt-7},{sin(160)*20pt-3}) and ({cos(-160)*20pt-7},{sin(-160)*20pt+3}) .. ({cos(-160)*20pt},{sin(-160)*20pt});
 \draw[very thick] ({cos(30)*20pt},{sin(30)*20pt}) .. controls ({cos(30)*20pt+14},{sin(30)*20pt-5}) and ({cos(-30)*20pt+14},{sin(-30)*20pt+5}) .. ({cos(-30)*20pt},{sin(-30)*20pt});
 \draw[very thick] ({cos(20)*20pt},{sin(20)*20pt}) .. controls ({cos(20)*20pt+7},{sin(20)*20pt-3}) and ({cos(-20)*20pt+7},{sin(-20)*20pt+3}) .. ({cos(-20)*20pt},{sin(-20)*20pt});
 \end{tikzpicture}
 &=\scalebox{0.25}{
 \begin{tikzpicture}[baseline=2]
 \tblockshortflip{0}{0}{0}
 \tblockshortflip{0}{12}{0}
 \tblockshortflip{0}{24}{0}
 \tblockshortflip{0}{36}{0}
 \wickU{4}{42}{7};
 \wickU{0}{46}{9};
 \wickU{6}{16}{5};
 \wickU{10}{12}{3};
 \wickU{18}{28}{5};
 \wickU{22}{24}{3};
 \wickU{18+12}{28+12}{5};
 \wickU{22+12}{24+12}{3};
 \draw[very thick] (-1,{\rt}) -- (-1,{-.25*11-1.5}) -- (47, {-.25*11-1.5}) -- (47, {\rt});
 \draw[thick] (11,{\rt}) circle (3pt);
 \filldraw[purple] (11,{\rt}) circle (3pt);
 \draw[thick] (23,{\rt}) circle (3pt);
 \filldraw[purple] (23,{\rt}) circle (3pt);
 \draw[thick] (35,{\rt}) circle (3pt);
 \filldraw[purple] (35,{\rt}) circle (3pt);
 \draw[thick] (35+12,{\rt}) circle (3pt);
 \filldraw[purple] (35+12,{\rt}) circle (3pt);
 \wickflipU{0}{22}{9}
 \wickflipU{4}{18}{7}
 \wickflipU{6}{16}{5}
 \wickflipU{10}{12}{3}
 \wickflipU{0+24}{22+24}{9}
 \wickflipU{4+24}{18+24}{7}
 \wickflipU{6+24}{16+24}{5}
 \wickflipU{10+24}{12+24}{3}
 \end{tikzpicture}}.
\end{align} 
The large $q$ limit ensures that for this second case of boundary conditions, only two-cycles contribute (so that all $\tau_i$ are identical permutations). This means we can look at all the $n$ lines as one, and so every quantum channel that appears will appear $n$ times
\begin{align}
 \begin{tikzpicture}[scale=1.5,baseline=-4]
 \centerarc[\darkblue,very thick](0,0)(0:45:20pt)
 \centerarc[\darkred,very thick](0,0)(45:90:20pt)
 \centerarc[\darkblue,very thick](0,0)(90:135:20pt)
 \centerarc[\darkred,very thick](0,0)(135:180:20pt)
 \centerarc[\darkblue,very thick](0,0)(180:225:20pt)
 \centerarc[\darkred,very thick](0,0)(225:270:20pt)
 \centerarc[\darkblue,very thick](0,0)(270:315:20pt)
 \centerarc[\darkred,very thick](0,0)(315:360:20pt)
 \filldraw[purple] (.7,0) circle (3pt);
 \filldraw[purple] (0,.7) circle (3pt);
 \filldraw[purple] (-.7,0) circle (3pt);
 \filldraw[purple] (0,-.7) circle (3pt);
 \filldraw[orange] ({.7*\rt},{.7*\rt}) circle (3pt);
 \filldraw[orange] ({-.7*\rt},{.7*\rt}) circle (3pt);
 \filldraw[orange] ({.7*\rt},{-.7*\rt}) circle (3pt);
 \filldraw[orange] ({-.7*\rt},{-.7*\rt}) circle (3pt);
 \draw[very thick] (.7,0) circle (3pt);
 \draw[very thick] (0,.7) circle (3pt);
 \draw[very thick] (-.7,0) circle (3pt);
 \draw[very thick] (0,-.7) circle (3pt);
 \draw[very thick] ({.7*\rt},{.7*\rt}) circle (3pt);
 \draw[very thick] ({-.7*\rt},{.7*\rt}) circle (3pt);
 \draw[very thick] ({.7*\rt},{-.7*\rt}) circle (3pt);
 \draw[very thick] ({-.7*\rt},{-.7*\rt}) circle (3pt);
 \draw[very thick] ({cos(75)*20pt},{sin(75)*20pt}) .. controls ({cos(75)*20pt-5},{sin(75)*20pt-5}) and ({cos(60+45)*20pt+5},{sin(60+45)*20pt-5}) .. ({cos(60+45)*20pt},{sin(60+45)*20pt});
 \draw[very thick] ({cos(60)*20pt},{sin(60)*20pt}) .. controls ({cos(60)*20pt-5},{sin(60)*20pt-5}) and ({cos(120)*20pt+5},{sin(120)*20pt-5}) .. ({cos(120)*20pt},{sin(120)*20pt});
 \draw[very thick] ({cos(75+180)*20pt},{sin(75+180)*20pt}) .. controls ({cos(75+180)*20pt+5},{sin(75+180)*20pt+5}) and ({cos(60+45+180)*20pt-5},{sin(60+45+180)*20pt+5}) .. ({cos(60+45+180)*20pt},{sin(60+45+180)*20pt});
 \draw[very thick] ({cos(60+180)*20pt},{sin(60+180)*20pt}) .. controls ({cos(60+180)*20pt+5},{sin(60+180)*20pt+5}) and ({cos(120+180)*20pt-5},{sin(120+180)*20pt+5}) .. ({cos(120+180)*20pt},{sin(120+180)*20pt});
 \draw[very thick] ({cos(150)*20pt},{sin(150)*20pt}) .. controls ({cos(150)*20pt-5},{sin(150)*20pt-5}) and ({cos(30)*20pt+5},{sin(30)*20pt-5}) .. ({cos(30)*20pt},{sin(30)*20pt});
 \draw[very thick] ({cos(160)*20pt},{sin(160)*20pt}) .. controls ({cos(160)*20pt-5},{sin(160)*20pt-5}) and ({cos(20)*20pt+5},{sin(20)*20pt-5}) .. ({cos(20)*20pt},{sin(20)*20pt});
 \draw[very thick] ({cos(-150)*20pt},{sin(-150)*20pt}) .. controls ({cos(-150)*20pt-5},{sin(-150)*20pt+5}) and ({cos(-30)*20pt-5},{sin(-30)*20pt+5}) .. ({cos(-30)*20pt},{sin(-30)*20pt});
 \draw[very thick] ({cos(-160)*20pt},{sin(-160)*20pt}) .. controls ({cos(-160)*20pt-5},{sin(-160)*20pt+5}) and ({cos(-20)*20pt-5},{sin(-20)*20pt+5}) .. ({cos(-20)*20pt},{sin(-20)*20pt});
 \node[scale = .3] at (0,14.5pt) {$\vdots$};
 \node[scale = .3] at (0,-13.5pt) {$\vdots$};
 \node[scale = .3] at (0,5pt) {$\vdots$};
 \node[scale = .3] at (0,-4pt) {$\vdots$};
 \end{tikzpicture}\quad \stackrel{q\rightarrow\infty}{\longrightarrow}\quad 
 \begin{tikzpicture}[scale=1.5,baseline=-4]
 \centerarc[\darkblue,very thick](0,0)(0:45:20pt)
 \centerarc[\darkred,very thick](0,0)(45:90:20pt)
 \centerarc[\darkblue,very thick](0,0)(90:135:20pt)
 \centerarc[\darkred,very thick](0,0)(135:180:20pt)
 \centerarc[\darkblue,very thick](0,0)(180:225:20pt)
 \centerarc[\darkred,very thick](0,0)(225:270:20pt)
 \centerarc[\darkblue,very thick](0,0)(270:315:20pt)
 \centerarc[\darkred,very thick](0,0)(315:360:20pt)
 \filldraw[purple] (.7,0) circle (3pt);
 \filldraw[purple] (0,.7) circle (3pt);
 \filldraw[purple] (-.7,0) circle (3pt);
 \filldraw[purple] (0,-.7) circle (3pt);
 \filldraw[orange] ({.7*\rt},{.7*\rt}) circle (3pt);
 \filldraw[orange] ({-.7*\rt},{.7*\rt}) circle (3pt);
 \filldraw[orange] ({.7*\rt},{-.7*\rt}) circle (3pt);
 \filldraw[orange] ({-.7*\rt},{-.7*\rt}) circle (3pt);
 \draw[very thick] (.7,0) circle (3pt);
 \draw[very thick] (0,.7) circle (3pt);
 \draw[very thick] (-.7,0) circle (3pt);
 \draw[very thick] (0,-.7) circle (3pt);
 \draw[very thick] ({.7*\rt},{.7*\rt}) circle (3pt);
 \draw[very thick] ({-.7*\rt},{.7*\rt}) circle (3pt);
 \draw[very thick] ({.7*\rt},{-.7*\rt}) circle (3pt);
 \draw[very thick] ({-.7*\rt},{-.7*\rt}) circle (3pt);
 \draw[line width=1mm] ({cos(60)*20pt},{sin(60)*20pt}) .. controls ({cos(60)*20pt-5},{sin(60)*20pt-5}) and ({cos(120)*20pt+5},{sin(120)*20pt-5}) .. ({cos(120)*20pt},{sin(120)*20pt});
 \draw[line width=1mm] ({cos(60+180)*20pt},{sin(60+180)*20pt}) .. controls ({cos(60+180)*20pt+5},{sin(60+180)*20pt+5}) and ({cos(120+180)*20pt-5},{sin(120+180)*20pt+5}) .. ({cos(120+180)*20pt},{sin(120+180)*20pt});
 \draw[line width=1mm] ({cos(150)*20pt},{sin(150)*20pt}) .. controls ({cos(150)*20pt-5},{sin(150)*20pt-5}) and ({cos(30)*20pt+5},{sin(30)*20pt-5}) .. ({cos(30)*20pt},{sin(30)*20pt});
 \draw[line width=1mm] ({cos(-150)*20pt},{sin(-150)*20pt}) .. controls ({cos(-150)*20pt-5},{sin(-150)*20pt+5}) and ({cos(-30)*20pt-5},{sin(-30)*20pt+5}) .. ({cos(-30)*20pt},{sin(-30)*20pt});
 \end{tikzpicture}
\end{align} 
We see that the action of the contraction of eigenvectors $\ket{\tau_1 \geq \dots \geq \tau_n}$ on the Pauli inputs ${\otimes}^k\sigma_\gamma$ can be expressed in terms of quantum channels of the form $\mathcal{N}_m^n$, with $m$ taking values in the cycle lengths in $K(\tau_i)$. These $\mathcal{N}_m^n$ channels are finally contracted according to the placements of cycles in the complement of the permutation to give a contribution to the OTOC. Because of the identity $|\tau| + |K(\tau)| = k+1$, we have that $|\tau|=k/2$ so that there are $|K(\tau)| = k/2+1$ many quantum channels. The maximal size of $m$ is $k/2$, so OTOCs to the power of $k$ are determined by the eigenvalues of $\mathcal{N}_1,\dots, \mathcal{N}_{k/2}$ 
\begin{align}
\begin{aligned}
 C_{\alpha\gamma}^{(k),(12)}(x,t)&=\frac{1}{q^{{kn}}} \sum_{\tau_i \in NC_k} \langle{L_\alpha^1} \ket{\tau_1 \geq \dots \geq \tau_n} \bra{\tau_1 \geq \dots \geq \tau_n} {R_\beta^2}\rangle\\
 &=\frac{1}{q}\sum_{\text{2-cycles $\tau\in NC_k$}} (\text{contractions of } \{\mathcal{N}^n_m(\sigma_\gamma \otimes\mathcal{P}_{\sigma_\gamma}^{{\otimes}^\frac{m-2}{2}} \otimes\sigma_\gamma)\}_{\text{m cycle length of $K(\tau)$}})
 \end{aligned}.
\end{align}
As $n$ becomes large, we have
\begin{align}
C_{\alpha\gamma}^{(k),(12)}(x,t)&=\frac{1}{q}\sum_{\text{2-cycles $\tau\in NC_k$}} \prod_{m \in cycles K(\tau)}\lambda_{m}^{ n}(\text{contractions of } \{\mathcal{N}_m\})\bra{\sigma_\gamma \otimes\mathcal{P}_{\sigma_\gamma}^{{\otimes}^\frac{m-2}{2}} \otimes\sigma_\gamma}\mathcal{O}_m\rangle,
\end{align}
where $\lambda_m$ is the maximal nontrivial eigenvalue of $\mathcal{N}_m$ with corresponding eigenoperator $\mathcal{O}_m$.
From \eqref{eq:2cyc} we know that there are $C_{k/2}$ permutations in the sum. Crucially, the number of terms in the sum is independent of $n$ and $m\leq k$ for a fixed $k$. Therefore, if each term exponentially decays with $n$, then the entire OTOC decays exponentially\footnote{One may have had concern that there are factors of $q$ in the sum that could potentially cause the OTOC to diverge at large $q$. This is easily seen to not be the case because the OTOC is bounded above by one via the Cauchy-Schwarz inequality
\begin{align}
C_{\alpha\gamma}^{(k)}(x,t)=\langle (\sigma_\alpha(0,0)\sigma_\gamma(x,t))^k\rangle \leq \sqrt{\langle \sigma_\alpha(0,0)^2 \rangle^k} \sqrt{\langle \sigma_\gamma(x,t)^2 \rangle^k} =1.
\end{align}
}.
The operator $\sigma_\gamma \otimes\mathcal{P}_{\sigma_\gamma}^{{\otimes}^\frac{m-2}{2}} \otimes\sigma_\gamma$ is traceless, so it has no support on the identity operator whose eigenvalue was one. We expect that, generically, there are no other eigenoperators with eigenvalue one, i.e.~$\lambda_m<1$. Hence, as $n$ becomes large, the OTOC exponentially decays with time. With knowledge of the precise unitary, $\lambda_m$ may be explicitly evaluated.

\subsubsection{Case III}
The third case of boundary conditions is
\begin{align}
\bra{L_2^{(k)}(\sigma_\alpha)} (T^{(k)}_{n})^{m} \ket{R_2^{(k)}(\sigma_\gamma)}.
\end{align}
The left boundary condition amounts to connecting lines symmetric with respect to the green points with rainbows of $n$ lines, and the right boundary condition amounts as before to connecting lines around the black points. Each contraction with $\ket{\tau_1 \geq \dots \geq \tau_n}$ has the product of $\{\mathcal{N}_m\}_m$ with $m$ corresponding to cycles in $\tau_i$ (composed from $\tau_n$ to $\tau_1$ starting from ${\otimes}^k\sigma_\gamma$) and the complements of $\tau_i$ (composed from $\tau_1$ to $\tau_n$). Not all $\tau_n$ needs to be the same because there are no constraints for them to be only composed of two cycles



\begin{align}
C_{\alpha\gamma}^{(k),(22)}(x,t) &\sim \sum_{\substack{\text{$\tau_i\in NC_k$}\\\tau_1 \geq \dots \geq \tau_n}} (\text{contractions of } ( (\underset{\substack{m_n \in\\ |cycles \\K(\tau_n)|}}{\otimes} \mathcal{N}_{m_n})\circ\dots\circ (\underset{\substack{m_1 \in\\ |cycles \\K(\tau_1)|}}{\otimes} \mathcal{N}_{m_1}(\sigma_\gamma \otimes\mathcal{P}_{\sigma_\gamma}^{{\otimes}^\frac{{m_1}-2}{2}} \otimes\sigma_\gamma)) 
 )).\end{align}

Where we are omitting some complicated combinations of SWAP and transpose operators between the compositions of each layer for the convenience of notation, but these do not change the fact that each layer is a channel nor introduce new unit eigenvectors, so do not alter the arguments. 



The difficulty comes from the fact that the sum over $n$-chains have a number of terms proportional to $n!$, and also that the number of terms in each term in the sum is proportional to $q^n$. Therefore, there is no clean upper bound available. We would find it surprising if this OTOC did not exponentially decay, even though we expect a conclusive answer would require a precise specification of the unitary operators involved, given that for $x-1$ and $x+1$ the OTOC decays exponentially and we would expect the OTOC to be a well-behaved function of x. Furthermore, the numerical results produce exponential decay without assuming any particular parity condition, suggesting that this behaviour is generic. 


\section{The Approach to Free Independence}

\label{sec:approacj}

In the previous section, we proved that all $2k$-point OTOCs decay exponentially with the distance from the light cone. This held for all Pauli operators, and so it is clear that holds for linear combinations of Pauli operators. Taking $\mathcal{A}_t$ to be the algebra of operators associated to a finite region at time $t$, we have thus proven that $\mathcal{A}_0$ and $\mathcal{A}_{T}$ exponentially approach free independence at large $T$. Moreover, there is pairwise free independence between all algebras of the form $\mathcal{A}_{n T}$ for $n \in \mathbb{Z}$. With this structure in hand, we may return to its implications using free harmonic analysis. This will serve as a benchmark for the following subsection, where we investigate early-time behavior.

A generalized Pauli operator, $\sigma_{\alpha}$, has eigenvalues that are evenly distributed at $\pm 1$ (the Bernoulli distribution)
\begin{align}
 \rho(\lambda) = \frac{\delta(\lambda - 1) +\delta(\lambda + 1)}{2}.
\end{align}
The Cauchy transform is thus
\begin{align}
 G(z) = \int d\lambda \rho(\lambda )\frac{1}{z-\lambda} = \frac{z}{z^2-1}
\end{align}
and the corresponding $\mathcal{R}$-transform is
\begin{align}
 \mathcal{R}(z) = \frac{\sqrt{1+4z^2}-1}{2z}.
\end{align}
By the additivity of $\mathcal{R}$-transforms
\begin{align}
 \lim_{T\rightarrow \infty} \mathcal{R}_{\sigma_{\alpha}(0)+\sigma_\gamma(T)}(z) = \frac{\sqrt{1+4z^2}-1}{z}.
\end{align}
Transforming back, the Cauchy transform is
\begin{align}
 \lim_{T\rightarrow \infty} G_{\sigma_{\alpha}(0)+\sigma_\gamma(T)}(z) = \frac{1}{\sqrt{z^2-4}},
\end{align}
from which we can read off the eigenvalue distribution
\begin{align}
 \lim_{T\rightarrow \infty} \rho_{\sigma_{\alpha}(0)+\sigma_\gamma(T)} (\lambda)= \frac{1}{\pi \sqrt{4-\lambda^2}} \delta_{\lambda\in [-2,2]},
\end{align}
the so-called \textit{arcsine} distribution.

We can continue evolving operators in time such that there is pairwise freeness. Adding $n$ freely independent operators and rescaling by $\sqrt{n}$, we have
\begin{align}
 \lim_{T\rightarrow \infty} \mathcal{R}_{\sigma_{\alpha}(0)+\dots + \sigma_\gamma(nT)} (z)= n \frac{1}{\sqrt{n}}\mathcal{R}(z/\sqrt{n}).
\end{align}
As $n$ becomes large, this approaches $\mathcal{R}_{\sigma_{\alpha}(0)+\dots + \sigma_\gamma(nT)} = z$ whose corresponding Cauchy transform is
\begin{align}
\lim_{T\rightarrow \infty} G_{\sigma_{\alpha}(0)+\dots + \sigma_\gamma(nT)}(z) = \frac{z -\sqrt{z^2-4}}{2}.
\end{align}
The corresponding eigenvalue distribution is then given by the semicircle distribution, the free analog of the Gaussian distribution
\begin{align}
 \lim_{T\rightarrow \infty} \rho_{\sigma_{\alpha}(0)+\dots + \sigma_\gamma(nT)}(\lambda) = \frac{\sqrt{4-\lambda^2}}{2\pi } \delta_{\lambda\in [-2,2]}.
\end{align}

\begin{figure}
 \centering
 \includegraphics[width=0.32\linewidth]{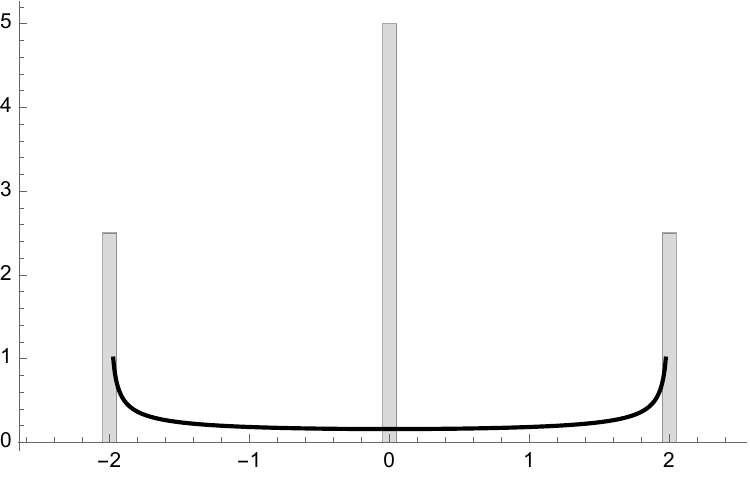}
 \includegraphics[width=0.32\linewidth]{spec_hp1_t0.pdf}
 \includegraphics[width=0.32\linewidth]{spec_hp1_t0.pdf}
 \includegraphics[width=0.32\linewidth]{spec_hp1_t0.pdf}
 \includegraphics[width=0.32\linewidth]{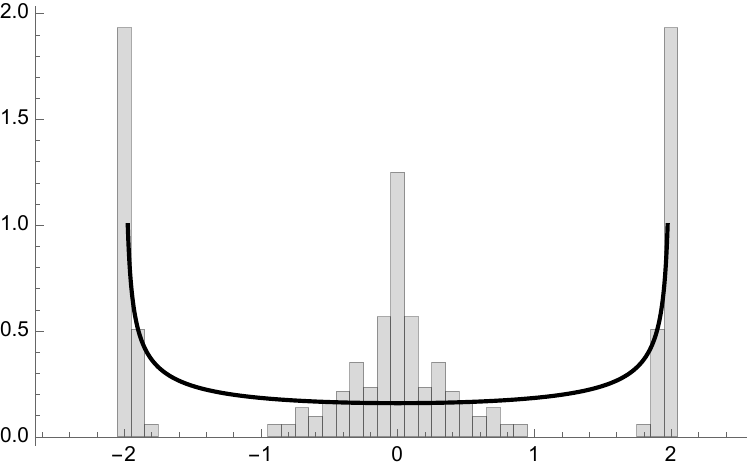}
 \includegraphics[width=0.32\linewidth]{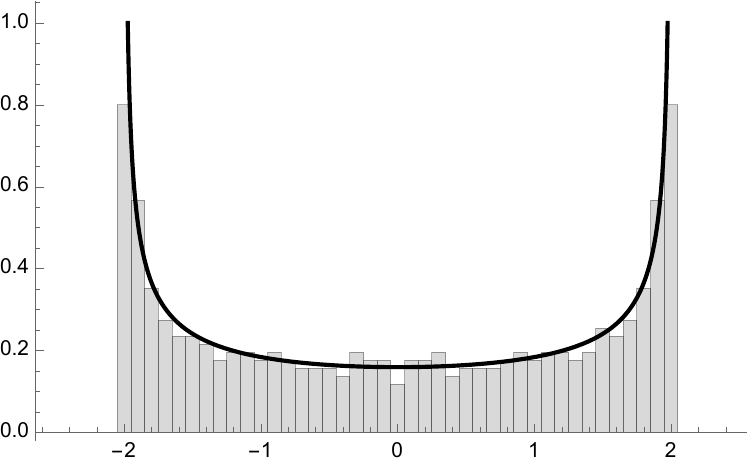}
 \includegraphics[width=0.32\linewidth]{spec_hp1_t0.pdf}
 \includegraphics[width=0.32\linewidth]{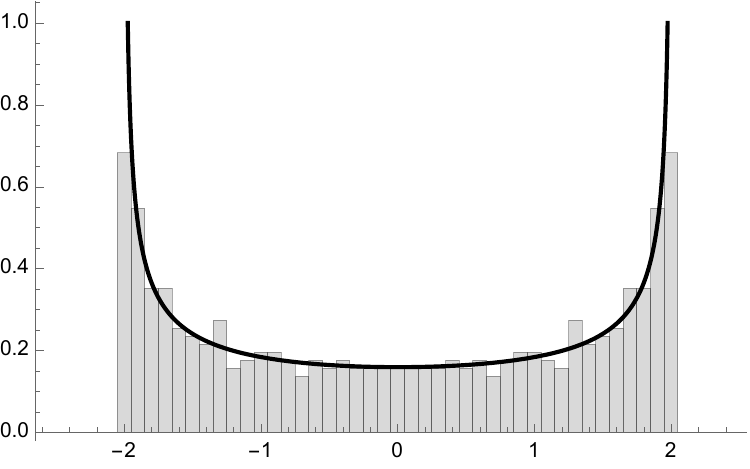}
 \includegraphics[width=0.32\linewidth]{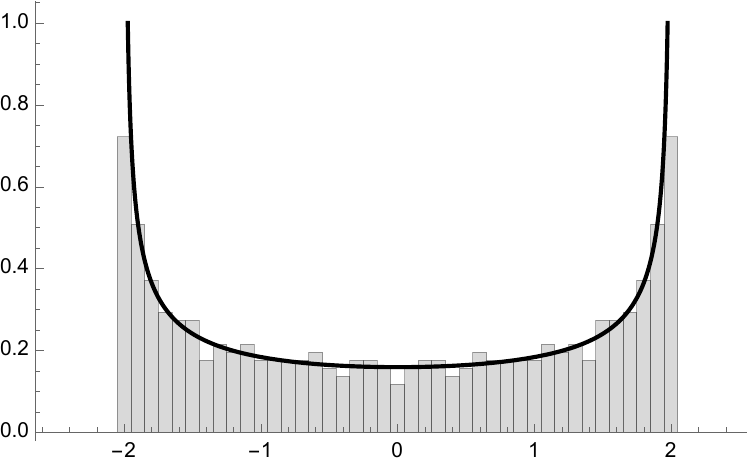}
 \caption{The spectra of $\sigma_x^{(4)} + \sigma_x^{(5)}(t)$ for the SDKI model at the integrable $h=0$ point (top row), near integrable $h = .1$ point (middle row) and chaotic $h = .5$ point (bottom row). The columns are for $t = 0,10,100$ respectively. The black line is the arcsine distribution, the prediction from free probability.}
 \label{fig:specs}
\end{figure}

\subsection{Self-Dual Kicked Ising}

\begin{figure}
\includegraphics[width = .32 \textwidth]{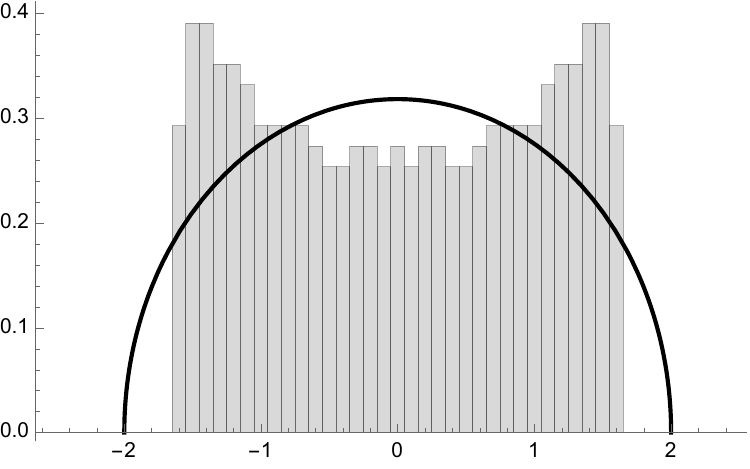}
\includegraphics[width = .32 \textwidth]{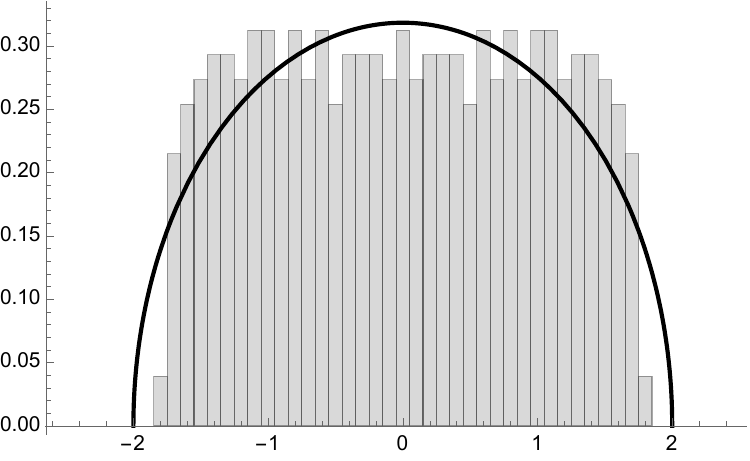}
\includegraphics[width = .32 \textwidth]{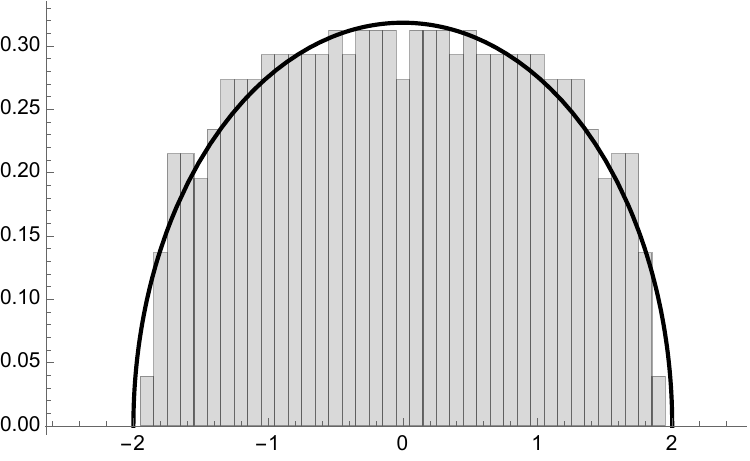}
\caption{The spectra of $\frac{1}{\sqrt{p_{max}}}\sum_{p=1}^{p_{max}} \sigma_x^{(4)}(10 p)$ for $p_{max} = 3$ (left), $5$ (middle), and $10$ (right) in the SDKI model at $h = 0.5$. The black line is the semicircle distribution.}
\label{semicirc}
\end{figure}

Thus far, we have provided analytic derivations of the OTOCs in the thermodynamic limit with $(x+t) \rightarrow \infty$. It is instructional to demonstrate the approach to free independence (or lack thereof) for finite size systems at finite times. For this purpose, we will numerically investigate the self-dual kicked Ising (SDKI) model, which is the canonical model exhibiting dual unitarity \cite{2018PhRvL.121z4101B,2016arXiv160207130A}. The Hamiltonian is given by
\begin{align}
\begin{aligned}
    H_{KI}[h;t] &= H_I[h] + H_K \sum_{m \in \mathbb{Z}}
    \delta (t - m\tau)
    \\
    H_I[h] &= \frac{\pi}{4\tau}\sum_{j}(\sigma_z^{(j)})\sigma^{(j+1)}_z+ h\sigma_z^{(j)}-\mathbbm{1}), \quad  H_K =  \frac{\pi}{4\tau}\sum_j\sigma_x^{(j)}
\end{aligned}
\end{align}
and the corresponding Floquet unitary is
\begin{align}
 U_{KI}[h] = e^{-iH_K}e^{-iH_I[h]} .
\end{align}
When $h = 0$, this model is integrable. Indeed, it is a Clifford circuit in this limit, mapping Pauli strings to Pauli strings. However, when $h$ takes a generic value, the model is chaotic. In Figure \ref{fig:specs}, we plot the spectra of the sum of two Pauli operators separated in time for different values of $h$ and compare the distribution to the arcsine distribution. At the integrable point, the spectrum does not change in time. It is always the classical convolution of two Bernoulli distributions. At a generic chaotic point, the spectrum rapidly approaches the arcsine distribution, demonstrating approximate free independence even in small systems. Close to the integrable point, we observe that the approach to approximate free independence is slower. We further demonstrate in Figure \ref{semicirc} that as more operators are summed, the distribution approaches the semicircle law.

\section{Discussion}
\label{sec:disc}
We conclude with a discussion of a few open questions and future directions.
\begin{itemize}
\item \textbf{Connecting to other aspects of deep thermalization} Deep thermalization was coined in the context of the ``projected ensemble,'' the notion that, in chaotic dynamics, ensembles of quantum states conditioned on measurements of the complement region become indistinguishable from the Haar ensemble up to the $k^{th}$ moment \cite{2023PRXQ....4a0311C, 2023Natur.613..468C}. Other closely related notions include the emergence of unitary $k$-designs \cite{2009CMaPh.291..257H, 2016CMaPh.346..397B, 2019arXiv190512053H}, late time growth of quantum complexity \cite{2018PhRvD..97h6015B,2018arXiv180202175S, 2022NatPh..18..528H}, higher order eigenstate thermalization \cite{2019PhRvE..99d2139F,2023arXiv230300713P}, and fluctuations in entanglement \cite{2020arXiv201011922C}. While clearly connected, it would be interesting to have a more precise characterization of what each notion implies about the others.
\item\textbf{Further application of replica trick} The replica trick is a remarkably useful method for computations in many-body physics, especially in relation to quantum information theory. Our explicit construction of the eigenstates of the replica transfer matrix should enable the calculation of many more interesting quantities, e.g.~such as operator entanglement and previously noted notions related to deep thermalization.
\item \textbf{Long wormholes and free product algebras in quantum gravity} Higher-order out-of-time-ordered correlators were considered for conformal field theories with holographic duals in \cite{2014JHEP...12..046S} (see also \cite{2018PhRvL.120l1601H}). Interestingly, the leading order bulk geometry describing these were found to be wormholes with very long interiors, growing with $k$. For these theories, the geometry explained the decay of the correlation functions at late times. This observation was leveraged in \cite{2023JHEP...04..009C} to construct free products of von Neumann algebras, relevant to understanding von Neumann entropy and the generalized second law of thermodynamics in quantum gravity. We believe an analogous free product construction may be enlightening in more general systems, such as dual-unitary circuits, where free independence of operators emerges at late times. The associated algebras that emerge here may be of particular interest mathematically, as they are type II$_1$ von Neumann algebras that are not hyperfinite.
\item \textbf{Quantifying freeness and the Lyapunov exponent} Without taking the thermodynamic limit, variables will never become truly freely independent. Nevertheless, we numerically found in Section \ref{sec:approacj} that the predictions from free probability theory are visible even for a small number of qubits at reasonably short times. It is desirable to quantify how ``close'' these operators are to being freely independent. One path may be to use distance measures, such as the statistical distance or Kullback-Leibler divergence, between the eigenvalue distributions and the distribution predicted from the free convolution. In certain systems with many degrees of freedom, there is an early time exponential behavior in the OTOC which has been interpreted as a quantum analog of a Lyapunov exponent \cite{2016JHEP...08..106M}. There is then some sense in which those systems exponentially approach free independence. It would be interesting if this could be quantified through measures of the distributions.
\item \textbf{Free entropy}
In free probability theory, there is a quantity called \textit{free entropy} that is analogous to Shannon entropy in classical probability theory \cite{2001math......3168V}. The theory of free entropy has led to progress in the understanding of type II$_{1}$ von Neumann algebras \cite{voiculescu1996analogues,ge1998applications,2004math......2108S,ge2002free}. While the definition of free entropy is quite involved, we note that it is additive for if and only if the operators are freely independent \cite{voiculescu1997free}. More generally, the entropy is subadditive. This suggests that the free mutual information may be a well-behaved measure of the extent to which variables are freely independent. Such a measure is generally challenging to evaluate, though it may be tractable for generalized Pauli matrices because the free entropy for projection operators are better understood \cite{2005math......4435H, 2006math......5633H}. 
\item \textbf{Relations between noncrossing partitions} Noncrossing partitions have come up in two seemingly distinct places in this work. The first was in the moment-cumulant formula for freely independent variables \eqref{eq:mom_cum}. The second was in classifying the eigenstates of the transfer matrix. It is not manifestly clear that these two instantiations are related to one another because free independence is expected for chaotic systems even without the dual-unitary property. Nevertheless, it is intriguing to examine the connections between the two.
\item \textbf{Second order freeness} Throughout this work, we have considered expectation values. Second order freeness is a generalization of free probability theory to understand fluctuations, moving beyond expectations \cite{2004math......5191M,2004math......5258M,2006math......6431C}. We expect this, even more detailed, structure to emerge at late times in chaotic quantum many-body systems. The combinatorial structure of second order freeness concerns noncrossing annular partitions and it would be interesting to understand if and how this structure arises in dual-unitary circuits.
\end{itemize}

\acknowledgments

We thank Giorgio Cipolloni, Austen Lamacraft, Silvia Pappalardi, Herman Verlinde, and Jinzhao Wang for valuable discussions and comments.
JKF is supported by the Marvin L.~Goldberger Member Fund at the Institute for Advanced Study and the National Science Foundation under Grant No. PHY-2207584. 

\bibliographystyle{JHEP.bst}
\bibliography{main}

\end{document}